\documentclass{aa} 

\usepackage{graphicx}
\usepackage{txfonts}
\usepackage{amsmath}

\begin{document}

\title{Comparative analysis of the influence of Sgr A* and nearby active galactic nuclei on the mass loss of known exoplanets}
\titlerunning{Influence of Sgr A* and nearby AGNs on the mass loss of exoplanets}

  \author{A. M. Wis\l{}ocka\inst{1}\fnmsep\inst{2},
     A. B. Kova\v{c}evi\'{c}\inst{2}
     \and
     A. Balbi\inst{3}
     }

  \institute{Institute for Astro- and Particle Physics, University of Innsbruck,
       Technikerstrasse 25/8, 620 Innsbruck, Austria\\
       \email{awismari@gmail.com}
     \and
       Department of Astronomy, Faculty of Mathematics, University of Belgrade, Studentski trg 16, 11000 Belgrade, Serbia\\
       \email{andjelka@matf.bg.ac.rs}
     \and
       Dipartimento di Fisica, Universit\`a degli Studi di Roma ``Tor Vergata'',
       Via della Ricerca Scientifica, I-00133 Rome, Italy \\
       \email{amedeo.balbi@roma2.infn.it}
       }

  \date{Received ..., 2018; accepted ..., 2019}


 \abstract
  {The detailed evolution of exoplanetary atmospheres has been the subject of decade-long studies. Only recently, investigations began on the possible atmospheric mass loss caused by the activity of galactic central engines. This question has so far been explored without using available exoplanet data.}
   {The goal of this paper is to improve our knowledge of the erosion of exoplanetary atmospheres through radiation from supermassive black holes (SMBHs) undergoing an active galactic nucleus (AGN) phase. 
   }
   {To this end, we extended the well-known energy-limited mass-loss model to include the case of radiation from AGNs. We set the fraction of incident power $\epsilon$ available to heat the atmosphere as either constant ($\epsilon = 0.1$) or flux dependent ($\epsilon = \epsilon(F_{\textrm{XUV}})$). We calculated the possible atmospheric mass loss for 54 known exoplanets (of which 16 are hot Jupiters residing in the Galactic bulge and 38 are Earth-like planets (EPs)) due to radiation from the Milky Way's (MW) central SMBH, Sagittarius A* (Sgr A*), and from a set of 107,220 AGNs generated using the 33,350 AGNs at $z < 0.5$ of the Sloan Digital Sky Survey database. }
   {We found that planets in the Galactic bulge might have lost up to several Earth atmospheres in mass during the AGN phase of Sgr A*, while the EPs are at a safe distance from Sgr A* ($> 7$ kpc) and have not undergone any atmospheric erosion in their lifetimes. We also found that the MW EPs might experience a mass loss up to $\sim 15$ times the Mars atmosphere over a period of $50$ Myr as the result of exposure to the cumulative extreme-UV flux $F_{\textrm{XUV}}$ from the AGNs up to $z = 0.5$. In both cases we found that an incorrect choice of $\epsilon$ can lead to significant mass loss overestimates.}
  {}


  \maketitle


\section{Introduction}

The evolution of exoplanetary atmospheres has been investigated in relation to the radiation from their host stars \citep[see, e.g.,][]{owen2015, bolmont2016, ribas2016}, as well as from nearby supernovae and gamma-ray bursts \citep{dartnell2011, melott2011}, but only recently has the possible influence of the Galactic central engine on their evolution come into focus \citep{balbi2017, forbes2018, chen2018}.

The supermassive black hole (SMBH) at the center of the Milky Way (MW), called Sagittarius A* (Sgr A*), has undergone a period of strong activity during its initial accretion phase, some 8 Gyr ago \citep{franceschini1999, ueda2014, brandt2015}. Observations also suggest more recent episodes of activity, for example, about 6 Myr ago \citep{nicastro2016}, and surprisingly even a century ago \citep{churazov2017}. Outbursts and jets from SMBHs have also been observed in many active galactic nuclei (AGNs), influencing both their host galaxies and the cosmic surroundings.

The available sample of known exoplanets is dominated by planets with volatile-rich atmospheres that are prone to erosion by ionizing radiation, extreme-UV (EUV) and X-rays (XUV). This provides strong motivation to investigate the long-term effect of the XUV irradiation by the neighboring AGNs on the atmospheres of known MW exoplanets. It is of particular importance to understand all the physical factors causing possible transition from gaseous to rocky planets. Moreover, a strong influx of ionizing radiation can affect the habitability of rocky exoplanets.

Exploration of the destructive potential of AGNs on galactic scales dates back to \citet{clarke1981} and \citet{gonzalez2005}, with the theoretical ground established by the pioneering works of \citet{chen2015}, \citet{balbi2017}, \citet{forbes2018} and \citet{chen2018}. For example, \citet{forbes2018} showed that the most important source of XUV radiation affecting planetary atmospheres except for a parent star is an accreting SMBH at the center of a galaxy. Additionally, \citet{chen2018} calculated that at distances $ \lesssim 100$ pc from a galactic center, the XUV flux at 1 AU for a G-type star is typically lower than that of an active SMBH. They also argued that the timescales of stellar activity, where the stellar XUV flux approaches that of an AGN, are disproportionately short compared to the duration of the AGN phase. From the work done so far on this topic, it becomes evident that even by itself, an AGN might influence planetary atmospheres to a large extent. All previous studies exploring the effect of AGNs on planetary atmospheres have focused purely on the relationship between the active phase of Sgr A* and either the Earth \citep{chen2015}, a generic sample of terrestrial planets \citep{balbi2017, forbes2018}, or sub-Neptunes \citep{chen2018} in the MW. In terms of the wider cosmological context, \citet{forbes2018} theoretically investigated the negative feedback of individual AGNs on the planets in the galaxies they are embedded in.

In our study we concentrate on the analysis of two unexplored questions in the context of the erosive effect of the XUV radiation from SMBHs on planetary atmospheres. The first is the effect of Sgr A* in its active and quiescent phases on known exoplanets in the MW. The second is an estimate of the cumulative feedback of XUV radiation of all AGNs up to $z = 0.5$. For this purpose, we use an extended model \citep{forbes2018, balbi2017} that is based on the study of stellar XUV radiation on planetary atmospheres.

Sgr A* has entered its active phase $\sim 8$ \textrm{Gyr} ago \citep{franceschini1999, ueda2014, brandt2015}, and it may have lasted between $\sim 10^{7-9}$ \textrm{yr} \citep{marconi2004}.
While the consideration of Sgr A* and its effect on the time evolution of planetary atmospheres is not new, studies of this topic so far have never examined this effect on existing exoplanet data (omitting Earth). To answer this question, we take full advantage of the open-access exoplanet database, the Extrasolar Planets Encyclopaedia\footnote{ http://exoplanet.eu/} (EPE), from which we extract data on 54 exoplanets. Our data set consists of 38 Earth-like planets (EPs) and 16 hot Jupiters residing in the Galactic bulge, found with the Sagittarius Window Eclipsing Extrasolar Planet Search (SWEEPS) mission \citep{sahu2007}, henceforth referred to as SWEEPS planets. Unlike previous studies \citep{chen2015, balbi2017, forbes2018, chen2018}, in addition to assuming the simplest, pure hydrogen (H) atmospheres for our planet sample, we also adapt the approach of \citet{bolmont2016} and address the mass-loss problem with hydrogen-oxygen (H/O) atmospheres (i.e., containing water (H$_2$O)).

We would like to emphasize the second question we address here with respect to the AGN demographic results. The Chandra mission data show that $\sim 10 - 20$ \% \citep{mushotzky2004} of all energy radiated over the lifetime of the Universe comes from AGNs, indicating that SMBHs must have influenced the formation of all structures in the Universe. Based on this observational evidence, the potential cumulative effect of the nearby AGNs on the atmospheres of MW planets should not be neglected. In addition, we perform a comparative analysis between the influence of Sgr A* and that of surrounding AGNs in search for clues to the exoplanet atmospheric mass loss over extended timescales.

We investigate the effect of SMBH on exoplanetary
 atmospheric evolution, excluding additional effects from the host stars. The atmospheric evolution of sub-Neptunian planets  at galactocentric distance of 10 pc has been investigated by  \cite{chen2018}, showing  that the XUV flux from a solar-like
 host star ($5.6$ ergs$^{-1}$cm$^{-2}$ at 1.0 AU and  850 ergs$^{-1}$cm$^{-2}$ at 0.1 AU)  is typically much lower than that from an SMBH ($4  \times 10^{4}$erg s$^{-1}$cm$^{-2}$ at 10 pc). At a distance of 1 AU the stellar radiation is about 150 times lower than at 0.1 AU from the host star, indicating an exponential decrease of radiation.

 If we assume a linear decrease of flux, then, at a distance 10 times  larger, a body receives 150 times lower radiation. This means that at about 2.5 AU, the stellar  radiation would be about $\sim$0.14 erg s$^{-1}$ cm$^{-2}$ (38 times lower than at 1 AU).
Moreover, these simulations show that periods of high stellar XUV  radiation, approaching that of the SMBH, only persist for a
 short time ($\sim 1000$ years).

 In our sample all exoplanets are within 8 kpc from the Galactic center. Even  if we assume the XUV flux that was obtained by \cite{chen2018}, the SMBH  XUV flux would be about 0.0625 erg s$^{-1}$  at 10 kpc, which is comparable  to what we obtained in our analysis. It is also very close to the limit proposed by \cite{bolmont2016},  who reported that planets are  affected by this radiation only when the XUV flux exceeds $\sim$ 0.1 erg s$^{-1}$ cm$^{-2}$.

In our study we do not investigate primary (proto)atmospheres of
planets. Primary atmospheres amount to only a few percent of  the total planetary mass, as in the case of e.g. Earth. In this case  \cite{Lehmer2017}    show
that  the saturated  flux phase of young stars (which can last for 100 Myr)  can predominantly affect  protoatmospheres of young planets.

The layout of this paper is as follows. In Section 2 we introduce the data sets we used in this study. We then outline our methods in Section 3. In Section 4 we present our results, and we discuss the implications of our work in Section 5, which is followed by our conclusions in Section 6.


\section{Data}
In order to quantify the impact of AGN radiation on exoplanetary atmospheres, we compile lists of available data on both types of objects, as described below.


\subsection{Exoplanet samples}
We included known EPs in our planet sample because future exoplanet projects are designed to characterize terrestrial exoplanet atmospheres as well as to search for atmospheric constituents considered as bio-markers (e.g., O$_3$, H$_2$O, and CH$_4$ \citep{oze2012, stv186, livio2017, schwieterman2018}). Such planets are of special interest because the study of atmospheric erosion in relation to the activity of SMBHs both in our galaxy and beyond it, with an emphasis on the latter, poses new questions regarding the concept of habitability.

Because the location within the Galaxy can play an important role in the evolution of planets, we also included the SWEEPS planets in our sample. To date, the SWEEPS planets, with distances $\sim 0.76$ kpc to Sgr A*, are the closest known exoplanets to the Galactic center. Moreover, all of the SWEEPS planets have previously been analyzed with respect to atmospheric mass loss through XUV fluxes from their parent stars \citep{sahu2006}. However, none were considered for a study of atmospheric erosion because of the XUV flux from an active or quiescent Srg A*, even though at such close vicinity it is quite likely that the AGN phase of the MW nucleus left a footprint on the mass content of their atmospheres.

Two important features of the Galactic environment need some attention. First, most of the Milky Way extinction curves are taken from stars in the solar neighborhood (within $\sim 2$ kpc), which only reflects the local properties of dust around the Sun. Extinction curves in the Galactic center and in the outer disk still remain uncertain because only a few, low-quality observations have been made. However, \cite{2017ApJ...840...27S} have shown that the dust associated with the Galactic center sightlines could also be considerably larger than that of the Galactic diffuse interstellar medium, indicating a smaller amount of extinction. Second, \cite{2004ApJS..152..211Z} and \cite{2013A&A...558A..62J}, modeling various dust species and grain size distributions, suggested that there are some different dust species that fit the Milky Way extinction curves, which can imply different extinction characteristics.

A comment is in order with regard to the orbital motion of planets. The solar galactocentric orbital motion has a period of $\sim~250$ Myr, while we considered an XUV irradiation over a time span of $\sim~50$ Myr. This time span is just $\sim~1/5$ of the Sun galactic orbit, which implies that the majority of earthlike planets in exoplanetary systems would only travel for small fractions of their galactic orbit during the considered period. If we assume that stars in the Milky Way bulge are moving around the Galactic center with velocities similar to our Sun, this means that stars with an orbital period of about 50 Myr will be about five times closer to Milky Way center than our Sun, which is on the order of the orbital motion of SWEEPS stars around the Galactic center. However, \cite{2017ApJ...840...27S} have shown that the dimensions of dust in the Galactic center could be larger than that of the Galactic diffuse interstellar medium (ISM), indicating that the extinction would be weaker. The effect of encountering ISM during the stellar galactic orbital motion is therefore not considered at our level of investigation.

To determine planetary atmospheric mass loss, specific planet parameters are required (see Section 3.1, Equation (1)). These are the distance to the irradiation source, which is found from coordinates, as well as mass and radius, which are used to determine density. For the loss of H/O atmospheres in the EP sample, additional information is needed: parent star radius and temperature to calculate the luminosity, and the semi-major axis of the planet to calculate the stellar XUV flux that reaches its atmosphere (Section 3.1.2).


\subsubsection{Sample of SWEEPS planets }

\begin{table*}
\caption{Data for the 16 SWEEPS planets.}
\label{table1}
\centering
\begin{tabular}{l l l l l l l }
\hline
Name & M (M$_\textrm{J}$) & M$_*$ (M$_{\odot}$) & R (R$_\textrm{J}$) & P (day) & a (AU) & d$_{\textrm{SgrA*}}$ (kpc) \\ \hline
SWEEPS-01 &     - & 0.81 &      1.01 &  1.56 &  0.025 & 0.76163 \\
SWEEPS-02 &     - &     0.55 &  1.37 &  0.912 & 0.015 & 0.76164 \\
SWEEPS-03 &     - &     0.79 &  0.87 &  1.27 &  0.021 & 0.76193 \\
SWEEPS-05 &     - &     0.66 &  1.09 &  2.313 & 0.03 & 0.76278 \\
SWEEPS-06 &     - &     1.09 &  0.82 &  3.039 & 0.042 & 0.76278 \\
SWEEPS-07 &     - &     1.09 &  0.9      &      1.747 & 0.027 & 0.76338  \\
SWEEPS-08 &     - &     0.87 &  0.98 &  0.868 & 0.017 & 0.76415 \\
SWEEPS-09 &     - &     0.79 &  1.01 &  1.617 & 0.025 & 0.76483 \\
SWEEPS-10 &     - &     0.44 &  1.24 & 0.424 &  0.008 & 0.76525 \\
SWEEPS-12 &     - &     0.86 &  0.91 & 2.952 &  0.038 & 0.76554 \\
SWEEPS-13 &     - &     0.91 &  0.78 & 1.684 &  0.027 & 0.76555 \\
SWEEPS-14 &     - &     0.8      & 0.93 & 2.965 &       0.037 & 0.76586 \\
SWEEPS-15 &     - &     0.49 & 1.37     & 0.541 &       0.01 & 0.76173  \\
SWEEPS-16 &     - & 0.68 &      1.4     & 0.969 &       0.017 & 0.76423 \\
SWEEPS-04 &     3.8 & 1.24 & 0.81 &     4.2     & 0.055  & 0.76193       \\
SWEEPS-11 &     9.7 & 1.1 &     1.13 &  1.796 & 0.03 & 0.76525 \\ \hline
\end{tabular}
\end{table*}

\begin{table*}
\caption{Excerpt of the table containing all the AGN data obtained from the SDSS database \citep{paris2018, abolfathi2018} in April 2018}
\label{table2}
\centering
\begin{tabular}{l l l l l l}
\hline
plate & fiberID & mid & z & ra (deg) & dec (deg) \\ \hline
        648 &   52559 &  142 & 0.000460623 &     356.5987 &     -10.39916 \\
        1114 &  53179 & 501     &       0.000460623 & 315.226 &  0.5804617\\
        7242 &  56628 & 262 &   0.000188514 &   24.12086 &      20.92283 \\
        845 &   52381 & 544 &   0.000460623 &   188.4821 &      5.718776 \\ \hline
\end{tabular}
\end{table*}

For our analysis we used the full SWEEPS sample, which includes 2 confirmed (SWEEPS $4$ and $11$) and 14 still unconfirmed planets \citep{sahu2007}. As shown in Table \ref{table1}, \citet{sahu2007} obtained the masses of SWEEPS $4$ and $11$. However, because the remaining 14 SWEEPS planets are unconfirmed, only their radii are determined. All of the SWEEPS planets are hot Jupiters with orbital periods $P < 10$ \textrm{days}, which enables us to estimate their masses with the mass-period (M-P) anticorrelation \citep[e.g.,][]{mazeh2005, henning2018} (see Appendix A.1 for a detailed discussion). In the sample, 5 of the planets (SWEEPS 2, 8, 10, 15, and 16) have even shorter orbital periods, $P < 1$ \textrm{day} \citep{sahu2007}, marking them as ultra-short hot Jupiters. These drastically short periods imply a runaway mass-transfer through the Roche-lobe overflow, as their orbits cross the Roche-limit of their stars \citep{valsecchi2014, valsecchi2015}. A period-mass-radius (P-M-R) relation exists for planets like this \citep{howell2001}, which we used to determine the masses of the ultra-short SWEEPS planets (see Appendix A.2 for details).

Here we would like to emphasize that regardless of the formation mechanism, tidal dissipation in the slowly spinning host stars causes the orbits of the tightest hot Jupiters currently known to shrink rapidly (e.g., \cite{1996Sci...274..954R, 2003ApJ...596.1327S, 2014MNRAS.440.1470B, 2014ApJ...787L...9V}, and references therein). Moreover, we note that the host stars of the tightest hot Jupiters are observed to currently rotate slowly. Eventually, hot Jupiters may decay down to their Roche-limit separation. Numerical simulations of the evolution of an exo-Jupiter with a 30 $M_{earth}$ show that for the first $\sim$ 70 Myr, tides cause the orbit to shrink, and the planet fills its Roche lobe \citep{valsecchi2015}. Valsecchi and collaborators also used the relation between period and radius of these planets from \cite{henning2018}

We also wish to point out that because some of the parameters of the SWEEPS planets are currently unknown, the above mass estimation methods are simply our best guess for the planet masses. For this reason, we explicitly included in our mass-loss calculations all of the estimates (see tables 4-6 and Figure 5). We also note that even if our estimates are not correct for the SWEEPS planets in question, the mass loss we obtain for them is not necessarily a meaningless result because it could be relevant for some arbitrary planet in the Galactic bulge with exactly these parameters.

The ages of all the SWEEPS planets are estimated to be $\sim 10$ Gyr \citep{sahu2007}, which indicates that they have lived through one of the most hostile events in the history of the MW, the AGN phase of Sgr A*. At that time, these planets would have been $\sim 2$ \textrm{Gyr} old, therefore it is expected that they already had fully developed atmospheres.


\subsubsection{Sample of Earth-like planets}

\begin{sidewaystable*}
\caption{ Data for 38 Earth-like planets, extracted in March 2018. The table also includes information on Earth because we include it in our calculations.}
\label{table3}
\centering
\begin{tabular}{l l l l l l l l l l l}
\hline
Name & M (M$_\textrm{J}$) & R (R$_\textrm{J}$) & $\rho$ (g cm$^{-3}$) & P (days) & M$_*$ (M$_{\odot}$) & Age (Gyrs) & T$_*$ (K) & d$_{\textrm{SgrA*}}$ (kpc) & ra (deg) & dec (deg) \\
\hline
Earth      & 0.003 & 0.089 & 5.50  & 365.26        & 1    & 4.6      & 5772  & 7.86         & - & -  \\
GJ 436 d     & 0.001 & 0.058 & 5.78  & -        & 0.452  & 6       & 3684  & 7.86        & 175.5458333 & 26.7063889  \\
GJ 436 c     & 0.001 & 0.059 & 5.81  & 1.37      & 0.452  & 6       & 3684  & 7.86        & 175.5458333 & 26.7063889  \\
TRAPPIST-1 d   & 0.001 & 0.069 & 5.27  & 4.05      & 0.08  & 0.5      & 2550  & 7.86        & 346.6208333 & -5.041388896 \\
TRAPPIST-1 e   & 0.002 & 0.082 & 4.83  & 6.1       & 0.08  & 0.5      & 2550  & 7.86        & 346.6208333 & -5.041388896 \\
Kepler-70 c   & 0.002 & 0.078 & 5.87  & 0.34      & 0.496  & 0.0184     & 27730  & 7.65        & 296.3541667 & 41.0927778  \\
TRAPPIST-1 f   & 0.002 & 0.093 & 3.44  & 9.21      & 0.08  & 0.5      & 2550  & 7.86        & 346.6208333 & -5.041388896 \\
Kepler-100 c   & 0.003 & 0.198 & 0.46  & 12.82      & 1.109  & 6.5      & 5825  & 7.78        & 291.3858333 & 41.99000003 \\
TRAPPIST-1 b   & 0.003 & 0.097 & 3.94  & 1.51      & 0.08  & 0.5      & 2550  & 7.86        & 346.6208333 & -5.041388896 \\
Kepler-109 b   & 0.004 & 0.209 & 0.60  & 6.48      & 1.069  & 6.35      & 5952  & 7.74        & 290.3927083 & 40.28486112 \\
TRAPPIST-1 g   & 0.004 & 0.101 & 5.51  & 12.35      & 0.08  & 0.5      & 2550  & 7.86        & 346.6208333 & -5.041388896 \\
TRAPPIST-1 c   & 0.004 & 0.094 & 6.88  & 2.42      & 0.08  & 0.5      & 2550  & 7.86        & 346.6208333 & -5.041388896 \\
GJ 1132 b    & 0.005 & 0.103 & 6.19  & 1.63      & 0.181  & 5       & 3270  & 7.86        & 153.7166667 & -47.15666668 \\
K2-3 c      & 0.007 & 0.165 & 1.95  & 24.64      & 0.612  & 1       & 3951  & 7.86        & 172.3333333 & -1.454722235 \\
Kepler-68 c   & 0.007 & 0.083 & 16.08  & 9.6       & 1.079  & 6.3      & 5793  & 7.84        & 291.0333333 & 49.04027778 \\
Kepler-109 c   & 0.007 & 0.235 & 0.71  & 21.22      & 1.069  & 6.35      & 5952  & 7.74        & 290.3927083 & 40.28486112 \\
Kepler-37 b   & 0.009 & 0.029 & 498.53 & 13.37      & 0.803  & 6       & 5417  & 7.84        & 284.59625  & 44.51817   \\
Kepler-100 d   & 0.009 & 0.135 & 5.06  & 35.33      & 1.109  & 6.5      & 5825  & 7.78        & 291.3858333 & 41.99000003 \\
Kepler-20 e   & 0.01 & 0.077 & 27.96  & 6.1       & 0.912  & 8.8      & 5466  & 7.79        & 287.7    & 42.3386111  \\
Kepler-10 b   & 0.01 & 0.131 & 6.12  & 0.84      & 0.91  & 10.6      & 5708  & 7.84        & 285.6791667 & 50.2413889  \\
Kepler-289 c   & 0.013 & 0.239 & 1.26  & 66.06      & 1.08  & 0.65      & 5990  & 7.74        & 297.4653321 & 42.882851  \\
Kepler-70 b   & 0.014 & 0.068 & 59.05  & 0.24      & 0.496  & 0.0184     & 27730  & 7.65        & 296.3541667 & 41.0927778  \\
CoRoT-7 b    & 0.015 & 0.136 & 7.86  & 0.85      & 0.93  & 1.5      & 5313  & 7.99        & 100.9541667 & -1.0627778  \\
HD 3167 b    & 0.016 & 0.152 & 5.97  & 0.96      & 0.88  & 0.83      & 5367  & 7.87        & 8.739666686 & 4.381472241 \\
Kepler-21 b   & 0.016 & 0.146 & 6.79  & 2.79      & 1.34  & 2.84      & 6131  & 7.82        & 287.3625  & 38.7138889  \\
K2-141 b     & 0.017 & 0.137 & 8.61  & 0.28      & 0.662  & 0.74      & 4373  & 7.85        & 350.9166667 & -1.189166675 \\
Kepler-92 c   & 0.018 & 0.219 & 2.26  & 26.72      & 1.209  & 5.52      & 5871  & 7.75        & 289.0875  & 41.5631   \\
CoRoT-24 b    & 0.018 & 0.33  & 0.66  & 5.11      & 0.91  & 11       & 4950  & 8.35        & 101.925   & -3.719444461 \\
K2-106 c     & 0.018 & 0.22  & 2.24  & 13.34      & 0.93  & 5       & 5814  & 7.95        & 13.0797796 & 10.7946987  \\
Kepler-68 b   & 0.019 & 0.206 & 2.85  & 5.4       & 1.079  & 6.3      & 5793  & 7.84        & 291.0333333 & 49.04027778 \\
GJ 1214 b    & 0.02 & 0.238 & 1.99  & 1.58      & 0.15  & 6       & 3026  & 7.85        & 258.8291667 & 4.9638889  \\
LHS 1140 b    & 0.021 & 0.128 & 13.22  & 24.74      & 0.146  & 5       & 3131  & 7.86        & 11.24583335 & -15.2713889 \\
Kepler-289 b   & 0.023 & 0.192 & 4.31  & 34.55      & 1.08  & 0.65      & 5990  & 7.74        & 297.4653321 & 42.882851  \\
Kepler-100 b   & 0.023 & 0.116 & 19.42  & 6.89      & 1.109  & 6.5      & 5825  & 7.78        & 291.3858333 & 41.99000003 \\
55 Cnc e     & 0.025 & 0.178 & 6.04  & 0.74      & 0.905  & 10.2      & 5196  & 7.87        & 133.1541667 & 28.3338889  \\
K2-3 b     & 0.026 & 0.194 & 4.72  & 10.05      & 0.612  & 1       & 3951  & 7.86        & 172.3333333 & -1.454722235 \\
K2-106 b     & 0.026 & 0.136 & 13.87  & 0.57      & 0.93  & 5       & 5814  & 7.95        & 13.0797796 & 10.7946987  \\
EPIC 210894022 b & 0.027 & 0.17  & 7.29  & 5.35      & 0.89  & 10.77     & 5730  & 8.05        & 59.88975418 & 21.29868613 \\
Kepler-88 b   & 0.027 & 0.337 & 0.95  & 10.95      & 1.022  & 2.45      & 5513  & 7.76        & 291.15   & 40.6694 \\
\hline
\end{tabular}
\end{sidewaystable*}

Table \ref{table3} displays data of the 38 EPs in our sample, which contains `subterran', `terran', and `superterran' planets ($0.1-0.5$ M$_{\oplus}$, $0.4 - 0.8$ R$_{\oplus}$; $0.5 - 5$ M$_{\oplus}$, $0.8 - 1.5$ R$_{\oplus}$; and $5 - 10$ M$_{\oplus}$, $1.5 - 2.5$ R$_{\oplus}$ \citep{stern2002}). We distinguish between the EPs in the sample with host star ages $> 8$ \textrm{Gyr} and $< 8$ \textrm{Gyr}, with the former already existing during the presumed activity period of Sgr A*. Throughout this text they are referred to as `older planets' and `younger planets', respectively. The assumption we make here is that the older planets underwent a period of enhanced XUV irradiation from the active Sgr A*, whereas the younger planets did not and their entire lifetime was spent bathed in radiation from its quiescent (current) phase.

Our EP sample contains five exoplanets that are believed to orbit within their stars' habitable zone: LHS 1140 b \citep{dittmann2017}, and TRAPPIST-1 d, e, f, and g \citep{gillon2016, delrez2018, turbet2018}. That these exoplanets are part of our sample presents an opportunity to asses the additional factors that might influence their long-term habitability.


\subsection{($z < 0.5$) AGN sample}
Because the radiative flux is inversely proportional to distance, $F \propto d^{-2}$, at first-order approximation, we expect only the closest AGNs to have an effect on the MW planets. We therefore chose  $z = 0.5$ as the limiting redshift. In Table \ref{table3} we show an excerpt of the data of the AGN sample obtained from the SDSS DR 14 database, which consists of $33,350$ AGNs up to $z = 0.5$ \citep{paris2018, abolfathi2018}. The total of SDSS observations cover only $\sim 1/3$ of the sky \citep{paris2018, abolfathi2018}. However, in this study we wish to investigate the effects of the global, full-sky distribution of the AGNs in the MW EPs. In Section 3.2 we describe how the AGNs observed by the SDSS were used to generate a full-sky AGN sample. The only information regarding the AGNs that is required for our calculations are redshifts and coordinates. The coordinates and $z$ are necessary to determine the luminosity distances of the AGNs to the EPs (see Section 3.1 and Appendix B). For simplicity we used Earth's position in the MW as a base point to calculate the distances of the EPs to every AGN. At the megaparsec scales involved in this investigation, a few parsec difference in distance are of little consequence.

\section{Methods}

\subsection{Atmospheric mass-loss}
Atmospheric evaporation resulting from AGN XUV irradiation should follow the energy-limited hydrodynamic escape mechanism \citep[see][]{balbi2017, forbes2018, chen2018}. The analytical solution to the total mass lost over a period of time $t_0$ is given by \citep{murray2009}

\begin{equation} \label{eq:mass_lost_total}
M_{\mathrm{lost}} = \frac{\epsilon F_{\mathrm{XUV}} \pi R_\mathrm{p}^2}{G M_\mathrm{p}/{R_\mathrm{p}}}t_\mathrm{0} \propto \frac{\epsilon}{\rho_\mathrm{p}} \frac{L_{\mathrm{XUV}}}{D_\mathrm{L}^2}t_\mathrm{0},
\end{equation}

\noindent where $R_\mathrm{p}$, $M_\mathrm{p}$ , and $\rho_\mathrm{p}$ are the planetary radius, mass, and density, respectively, $D_\mathrm{L}$ is the luminosity distance\footnote{ The definition of $D_\mathrm{L}$ changes on cosmological scales, therefore it is calculated differently for Sgr A* as compared to the AGNs at $z < 0.5$. In our calculations we assumed the following cosmology: $H_\mathrm{0} = 73$ \textrm{km s$^{-1}$ Mpc$^{-1}$}, $\Omega_\mathrm{m} = 0.27$, $\Omega_\Lambda = 0.73$. See Appendix B for an expanded explanation.} to the source, and $G$ is the gravitational constant. The XUV luminosity is given by $L_{\mathrm{XUV}} = \eta_\mathrm{X} L_{\mathrm{bol}}$, where $\eta_\mathrm{X}$ is the high-energy photon fraction of bolometric luminosity $L_{\mathrm{bol}}$ of a source. We chose $\eta_\mathrm{X} = 0.7$ \citep{vasudevan2009, lusso2012, balbi2017} for all AGNs and also for the quiescent period of SgrA*. In the energy-limited regime the uncertainties in heating and cooling are included in the efficiency parameter $\epsilon$ (Section 3.1.1). Interestingly, the total mass loss is $\propto \rho_\mathrm{p}^{-1}$ in this model in terms of planetary parameters.

\citet{murray2009} found that the analytical solution from Equation (\ref{eq:mass_lost_total}) is close to the result found with complex numerical calculations, implemented into codes such as ZEUS \citep{stone1992, hayes2006, owen2015, bolmont2016} or PLUTO-CLOUDY \citep{salz2015}. We therefore implemented the analytical approach in our work.

The energy-limited solution is applicable for XUV fluxes in the range $0.1 \leqslant F_{\mathrm{XUV}} \lesssim 10^{3-4}$ \textrm{erg s$^{-1}$ cm$^{-2}$} \citep{murray2009, bolmont2016}, otherwise the flow enters a different regime and has to be treated accordingly \citep{murray2009, owen2015}. In Section 4 we show that for any of the planets in our sample, the $F_{\mathrm{XUV}}$ from the AGNs either decreases within the energy-limited range or it is too low to induce any wind. In the case of the H/O atmospheres, the $F_{\mathrm{XUV}}$ rises to the point when the radiation-limited regime is entered (see Section 4.2.2).

To solve Equation (\ref{eq:mass_lost_total}), we require the $L_{\mathrm{XUV}}$ of Sgr A* and all local AGNs. A simplifying assumption made in this work is a quasar duty cycle for all AGNs for a timescale on the order of the Salpeter time, that is, the mass-independent time it takes to double the mass of a black hole in the Eddington accretion phase \citep{forbes2018}. We estimated the Eddington luminosities of the AGNs with a black hole mass $M_{\mathrm{BH}}$ using the standard definition, $L_{\mathrm{Edd}} \approx 1.3 \times 10^{38} (M_{\mathrm{BH}}/M_\odot)$ \textrm{erg s$^{-1}$}. Taking the mass of Sgr A*, $M_{\mathrm{BH}}= 3.6 (\pm 0.3) \times10^6 M_{\odot}$ \citep{schodel2009}, the resulting Eddington luminosity is $L_{\mathrm{Edd}} \approx 4.7 \times 10^{44}$ \textrm{erg s$^{-1}$}, while in its current quiescent state, its measured luminosity is surprisingly low at $L_{\textrm{Sgr A*}} \simeq 10^{-8.5} L_{\textrm{Edd}}$ \citep{sabha2010}. For the sample of the AGNs with $z < 0.5,$ we assumed black hole masses in the range $M_{\mathrm{BH}} = 10^{7-9} M_{\odot}$ \citep[see][Figure 1]{mclure2004}, with which we obtained luminosity ranges of $L_{\mathrm{Edd}} = 10^{45-47}$ \textrm{erg s$^{-1}$}. We obtained a value of $\bigtriangleup t_\mathrm{S}= M_{\mathrm{BH}} / \dot{M}_{\mathrm{BH}} \approx 50$ \textrm{Myr} for the Salpeter time of Sgr A*, where $\dot{M}_{BH} = (1- \epsilon_{\mathrm{BH}})\dot{M}_{\mathrm{acc}}$ is the rate of mass gained by the black hole with mass accretion rate $\dot{M}_{\mathrm{acc}}$ and radiative efficiency $\epsilon_{\mathrm{BH}}$, which has a typical value of $0.1$. We used $t_\mathrm{0} = \bigtriangleup t_\mathrm{S}$ in Equation (\ref{eq:mass_lost_total}), and we assumed the same accretion timescale as calculated for Sgr A* for all AGNs in our sample, where we chose to bypass the time dilation effect applicable on cosmological scales because the $z$ of these AGNs is small enough to avoid causing a significant alteration to $t_\mathrm{0}$ (see Appendix C for a thorough explanation). The assumption of $\bigtriangleup t_\mathrm{S} \approx 50$ Myr is consistent with numerous models exploring the $\bigtriangleup t_\mathrm{S}$ in the past decades \citep{richstone1998, haehnelt1998, martini2001, hopkins2005}.

We note that we assume no photon losses throughout, that is, an opacity $\tau = 0$. This assumption clearly maximizes the atmospheric mass loss that we calculate for our sample of planets. We comment on this point in Section 3.2 in greater details.

\subsubsection{Efficiency parameter $\epsilon$ }

A common approach in energy-limited escape calculations is to use a constant value of $\epsilon = 0.1$ \citep[e.g.,][]{watson1981, lopez2012, des2012}. \citet{bolmont2016} argued that in the wrong circumstances, this assumption can lead to an overestimated atmospheric mass loss. Likewise, \citet{owen2013} demonstrated that the efficiency parameter changes with mass and radius of the planet and with the incoming XUV flux. They also indicated that $\epsilon$ is varies throughout planetary evolution. We left this detail aside in our investigation because the mass-loss timescale we are interested in is far too short to require including the evolution of $\epsilon$ in our calculations. The assumption of H-only atmospheres also greatly simplifies the problem of atmospheric loss for various reasons. Because heating or cooling is included in the $\epsilon$ parameter, the wind efficiency is inevitably affected by complex particles. The presence of heavier elements or molecules results in radiative and collisional cooling of the gas \citep{owen2012, owen2013, bolmont2016}.

\citet{bolmont2016} used the models of \citet{owen2015} to calculate $\epsilon$ as a function of the incoming $F_{\textrm{XUV}}$, that is, $\epsilon = \epsilon(F_{\textrm{XUV}})$, for a H-only atmosphere. We refer to Figure 2 of their paper, from which it is inferred that at higher fluxes, $F_{\textrm{XUV}} \gtrsim 0.2$ \textrm{erg s$^{-1}$ cm$^{-2}$}, $\epsilon$ reaches its maximum value of $0.1$, where it remains in the energy-limited regime, until it begins to drop again at $F_{\textrm{XUV}} \approx 450$ \textrm{erg s$^{-1}$ cm$^{-2}$}. At higher fluxes, $\epsilon$ drops because radiative cooling increases and the wind reaches the recombination-limited regime \citep{murray2009}. Conversely, at lower fluxes, the hydrodynamical wind is never launched because the heating rate from the incoming XUV photons is simply insufficient \citep{bolmont2016}. There exists a minimum value of the incoming $F_{\textrm{XUV}}$, $F_{\textrm{min}} = 0.1$ \textrm{erg s$^{-1}$ cm$^{-2}$}, at which a wind can still be induced. At this value of XUV flux, $\epsilon \approx 0.02$ \citep[see][Figure 2]{bolmont2016}. Therefore, within the range $0.02 \lesssim \epsilon \leqslant 0.1$ the wind can be considered as a less powerful in the energy-limited regime.

\subsubsection{Hydrogen-oxygen atmospheres}

In the investigation of the effects of XUV irradiation on H/O atmospheres, we adopted the method suggested by \citet{bolmont2016} to explore the joint escape of H and O atoms. When there are O atoms present in the atmosphere, there is a chance that they will be dragged along with the H atom flux out of the planet’s potential. We assumed that the atmospheric content of H to O in the atmosphere remains the same at a 2:1 ratio. This implies an enormous water reservoir on the planet because this approach does not account for the evolution of the H/O ratio \citep{bolmont2016}. We note here that this assumption may be realistic. As an example, wer refer to the careful analysis of Kepler data, which revealed a bimodal planet distribution. One of these modes is related to water worlds, presumably made up of ($> 25 \% $) water \citep{zeng2018}.

In this approach, the mass-loss flux $F_\mathrm{M}$ is equivalent to the total mass-loss $M_{\mathrm{lost}}$ divided by the time of exposure to XUV radiation $t_\mathrm{0}$ and the cross-sectional area of the planet $4 \pi R_\mathrm{p}^2$, as follows:

\begin{equation} \label{eq:mass_flux3}
F_\textrm{M} = \frac{M_{\mathrm{lost}}}{4 \pi t_\mathrm{0} R_\textrm{p}^2} = \frac{3\epsilon F_{\textrm{XUV}}}{16 \pi t_\mathrm{0} G R_\textrm{p}^2 \rho_\textrm{p}},
\end{equation}

\noindent where we kept $\epsilon = 0.1$. Next, the ratio $r_\mathrm{F}$ for the escape fluxes of the particles, $F_\mathrm{O}$ and $F_\mathrm{H}$, in a hydrodynamic outflow, provided by \cite{hunten1987},

\begin{equation} \label{eq:flux_fraction}
r_\textrm{F} = \frac{F_\textrm{O}}{F_\textrm{H}} = \frac{X_\textrm{O} m_\textrm{c} - m_\textrm{O}}{X_\textrm{H} m_\textrm{c} - m_\textrm{H}},
\end{equation}

\noindent where $X_\mathrm{O}$ and $X_\mathrm{H}$ are the fractional abundances of the O and H atoms, which we assumed to be $X_\mathrm{O} = \frac{1}{3}$ and $X_\mathrm{H}=\frac{2}{3}$ (proportion of dissociated water), $m_\textrm{H}$ and $m_\textrm{O}$ are the mass of a single H and O atom and $m_\mathrm{c}$ is the cross-over mass, defined as

\begin{equation} \label{eq:m_c}
m_\mathrm{c} = m_\mathrm{H} + \frac{k_\mathrm{b} T F_\mathrm{H}}{bg X_\mathrm{H}},
\end{equation}

\noindent where $g$ is the planetary gravity, $k_\textrm{b}$ is the Boltzmann constant, and $b = 2.2 \times 10^{19}$ \textrm{cm$^{-1}$ s$^{-1}$} is the diffusion parameter \citep{hunten1987}. $T$ is the temperature of the wind, generally found with 1D radiation-hydrodynamic mass-loss simulations \citep{murray2009, owen2015, bolmont2016}. For exoplanets in close orbit of their star, the range of wind temperatures is found to lie between $3000 - 10^4$ \textrm{K} \citep{lammer2003a, murray2009, owen2012, bolmont2016}.

In order to use Equations \ref{eq:mass_flux3} and \ref{eq:flux_fraction}, we need to infer $F_\textrm{H}$. \citet{bolmont2016} defined the mass flux of H atoms $F_\textrm{H}$ as

\begin{equation} \label{eq:mass_fluxH}
F_\textrm{H} = \frac{F_\textrm{M} + m_\textrm{O} X_\textrm{O}(m_\textrm{O} - m_\textrm{H})\frac{bg}{k_\textrm{b} T}}{m_\textrm{H} + m_\textrm{O}\frac{X_\textrm{O}}{X_\textrm{H}}}
,\end{equation}

\noindent where for the O atoms to be dragged along with the wind, the condition $m_\mathrm{c} > m_\textrm{O}$ must be fulfilled, otherwise, when $m_\textrm{c} < m_\textrm{O}$ , only H atoms escape \citep{bolmont2016}.

To solve for $F_\textrm{H}$ , we need to know the temperature of the wind $T$. We estimated $T$ as the temperature due to the XUV radiation $T_{\mathrm{XUV}}$ reaching the planetary exosphere. As the additive nature of flux indicates, photons from Sgr A* would contribute to heating of the upper atmosphere and aid in mass loss. Therefore, the XUV flux is that of the planet’s parent star in addition to the XUV flux reaching it from Sgr A*. The scaling relation used in order to estimate the wind temperature of the planet $T_{\mathrm{XUV}, p}$ is adapted from \citet{lammer2003a},

\begin{equation} \label{eq:T_xuv}
T_{\textrm{XUV, p}}^s = \frac{(T_{\textrm{XUV}}^s - T_{\textrm{BB}}^s)_\textrm{e} F_{\textrm{XUV, p}}g_\textrm{e}}{F_{\textrm{XUV, e}}g_\textrm{p}} + T_{\textrm{BB, p}}^s,\end{equation}

\noindent where $s=0.7$ for neutral gas, $g$ is the acceleration due to gravitation, and the subscripts \textrm{e} and \textrm{p} indicate the parameters of Earth and the planet, respectively. The $T_{\textrm{BB}}$ is the blackbody (BB) temperature, which we calculated for every planet from the combined bolometric flux of the parent star and Sgr A* with

\begin{equation} \label{eq:T_eff}
T_{\textrm{BB}} = \left[\frac{F_*(1-A)}{4\sigma}\right]^{1/4},
\end{equation}

\noindent where $\sigma$ is the Stefan-Boltzman constant and $A$ is the planet albedo, which for simplicity, we assumed to be 0. According to observations, most hot Jupiters are quite dark at optical wavelengths. \cite{2008ApJ...689.1345R} used MOST data and derived a geometric albedo of 0.038 $\pm$ 0.045 for HD 209458b. \cite{2010A&A...513A..76S} measured 0.164 $\pm$ 0.032 for CoRoT-2b, but noted that most of this flux likely arose from optical thermal emission, not from scattered light. \cite{2012ApJ...761...53B} found based on Kepler data 0.0136 $\pm$ 0.0027 for TrES-2b, currently the ``darkest'' planet. The bulk of the evidence is that clouds do not have a dominant role in absorption or scattering of stellar light for this class of planets. However, there are outliers. In particular, Kepler-7b was found to have an albedo 0.35$\pm$ 0.02, likely indicating high silicate clouds \citep{2011ApJ...735L..12D}. Earthlike planets with a water-covered surface and those that are completely free of ice can be assumed to have 0.07, while the measured albedo of Earth's surface is about 0.13 \citep{2016A&A...592A..36G}. The planetary physical parametric space is huge and not well understood, therefore we limit it to what we know from recent observations and theoretical numerical investigations.

Equation (\ref{eq:T_xuv}) relates the $T_{\textrm{XUV}}$ of two planets of comparable gas compositions (we assumed H/O atmospheres for all EPs) to the XUV flux $F_{\mathrm{XUV, p}}$ reaching their atmospheres. To determine the wind temperature of the EPs, we assumed that the wind temperature of Earth is equivalent to the minimum temperature $T_{\mathrm{min}}$ required for an H atom to escape its gravitational potential (\citet{owen2015uv}), that is, $T_{\textrm{XUV, e}} = T_{\mathrm{min}}$, which we find to be $T_{\mathrm{min}} = 5035$ K, with

\begin{equation} \label{eq:T_escape}
T_{\textrm{escape}} = \frac{2}{3} \frac{G M_\textrm{p} m_\textrm{H}}{k_\textrm{b} R_\textrm{p}},
\end{equation}

\begin{figure}
\centering
\includegraphics[width=\hsize]{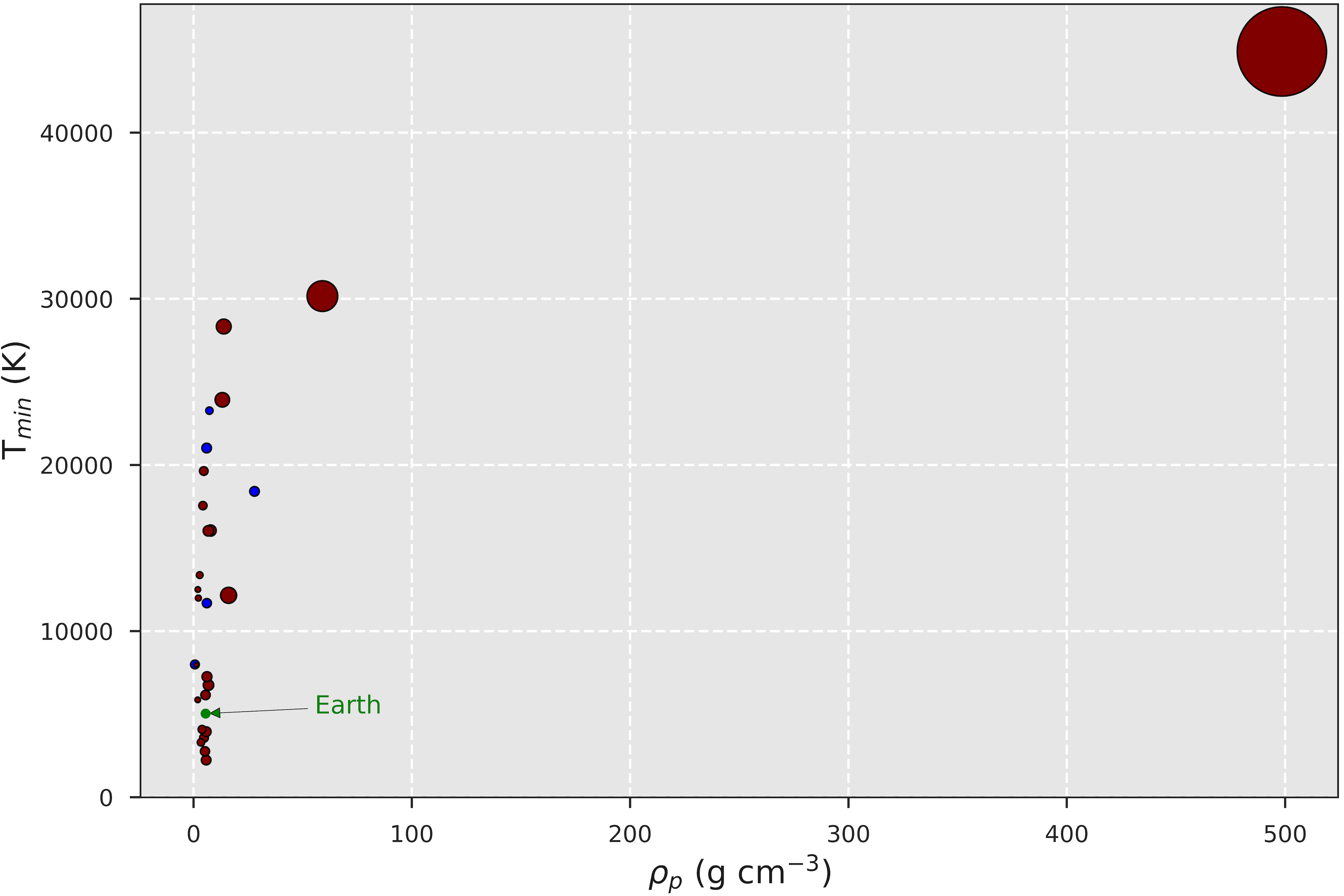}
\includegraphics[width=\hsize]{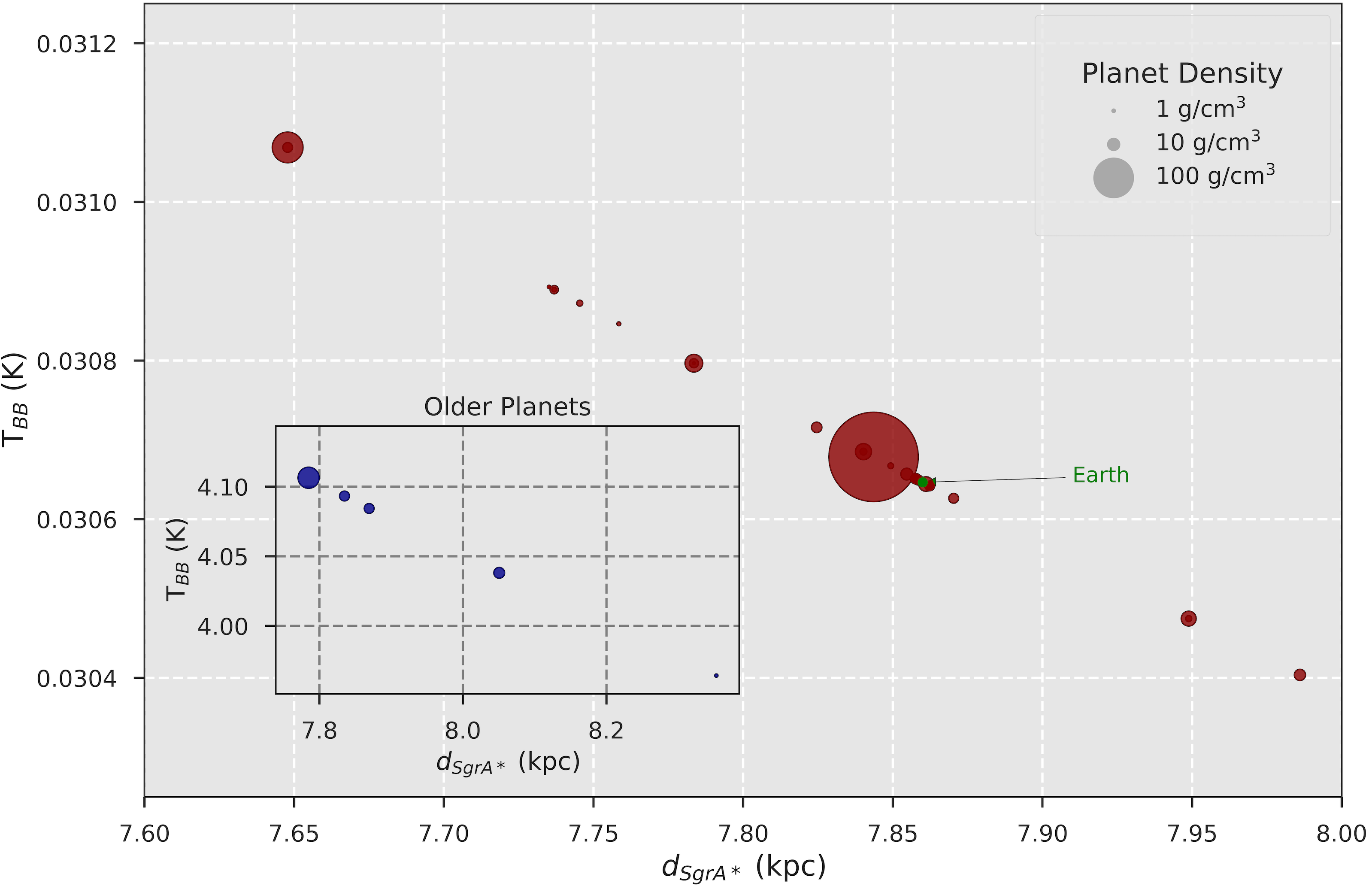}
\caption{(Top) Thermal energy required for an H atom to escape a planetary potential as a function of planet density. The size of each data point reflects the density of the planet. For example, for Earth, $T_{\textrm{min}} = 5035$ K, whereas the denser planets have much higher tempratures, above $10000$ K. (Bottom) BB temperature vs. distance from Sgr A* of the 28 EPs, purely due to heating from the active (inserted subplot: older EPs) and the quiescent (main plot: younger EPs) states of Sgr A*.}
\label{Tmin_and_BB}
\end{figure}

\begin{figure}
\centering
\includegraphics[width=\hsize]{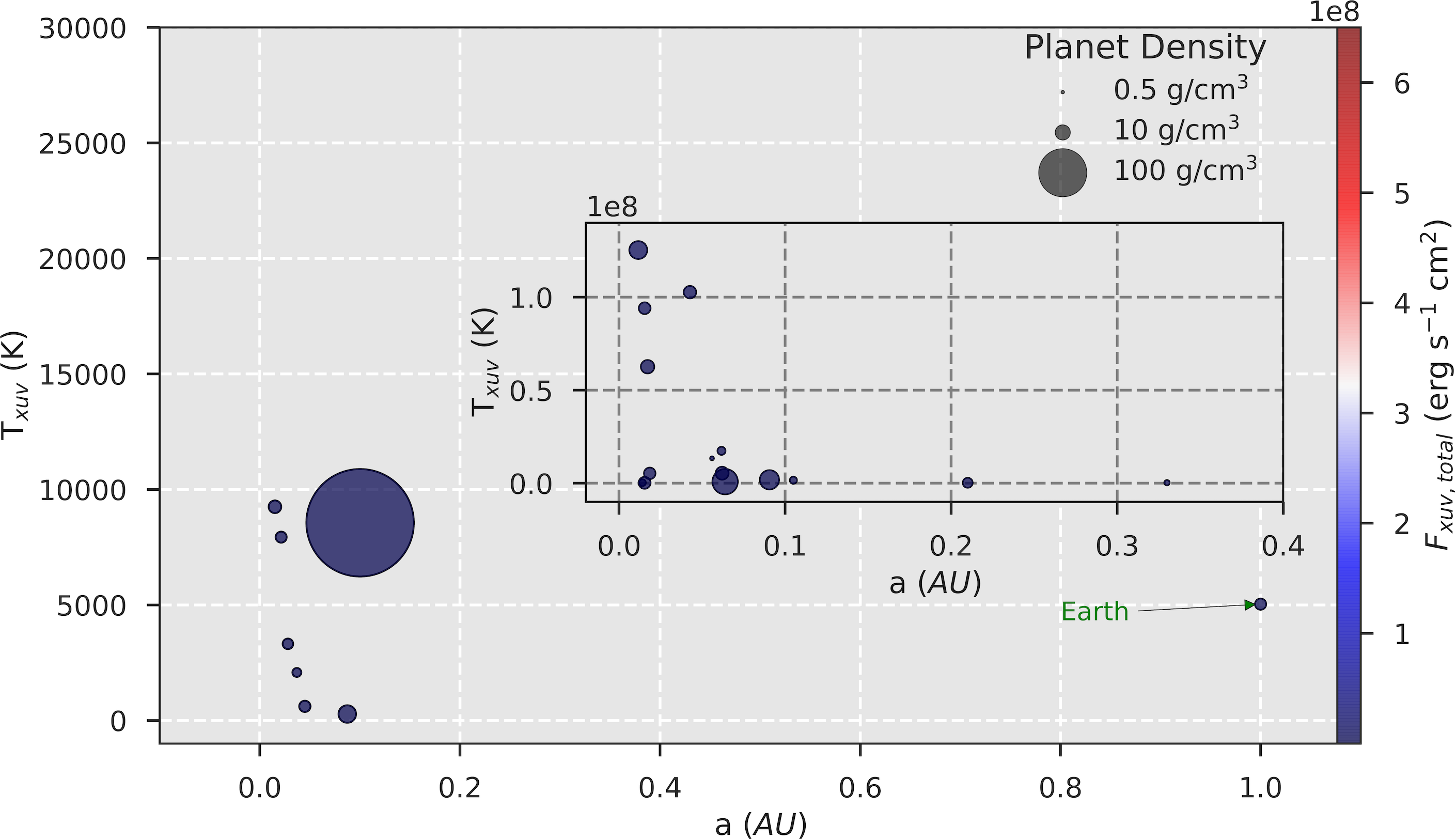}
\caption{$T_{\textrm{XUV}}$ of the 28 EPs as the result of heating by the stellar and Sgr A* XUV flux. Some of the planets are not represented in the plot for clarity.}
\label{Txuv}
\end{figure}

\noindent where $M_\textrm{p}$ and $R_\textrm{p}$ are the planet mass and radius, respectively. In Figure \ref{Tmin_and_BB} (top) we show $T_{\mathrm{min}}$ of all the EPs in our sample. The blue and red points represent the older and younger planets, respectively.

For this analysis our EP sample is reduced to 28 exoplanets because in addition to the other stellar parameters we need for our calculations, we also required the stellar luminosity to obtain the $T_{\textrm{BB}}$ of the planets (Section 2.1). Here we assumed that all stars lie on the main sequence (MS) and have $\eta_\mathrm{X} = 10^{-3}$. Figure \ref{Tmin_and_BB} (bottom) shows the BB temperature of the EPs purely due to the bolometric flux of Sgr A* as a function of distance $d_{\textrm{Sgr A*}}$ to Sgr A*. The inserted subplot gives the BB temperature ranges for the older planets, which encountered the flux of the active Sgr A*, hence the higher temperatures, $\sim 4$ \textrm{K}. The main plot shows the BB temperatures of the younger EPs, which only experienced flux of the quiescent Sgr A*, which results in much lower temperatures, $\sim 0.03$ \textrm{K}. Although the temperatures do not simply add up, this plot is included as a demonstration of the weak influence that Sgr A* has on the heating of atmospheres compared to the host stars of the planets.

The calculated $T_{\textrm{XUV, p}}$ for each EP is displayed in Figure \ref{Txuv}, where not all of the 28 planets from the EP sample are plotted for clarity. The XUV flux from Sgr A* contributes a fraction of the total XUV flux traversing the planets: $0.039-0.045$ \textrm{erg s$^{-1}$ cm$^{-2}$} and $1.36 \times 10^{-10} - 1.48 \times 10^{-10}$ \textrm{erg s$^{-1}$ cm$^{-2}$} for the older and younger planets, respectively. The figure shows that the wind temperatures of 7 of the planets (main plot) reach $10^3 - 10^4$ K, while another 11 planets (insert subplot) have significantly higher $T_{\textrm{XUV, p}}$, in the range $10^7 - 10^{11}$ \textrm{K}. The reason that these soaring temperatures result from our calculation is mainly related to three factors: 1) the orbital distance of the planet around its star (Kepler-70 c, Kepler-70 b, Kepler-21 b, Kepler-68 b, CoRoT-7 b, GJ 436 c, and K2-106 b; see Table 3.2), 2) the parent star temperature (Kepler-70 c and Kepler-70 b), and 3) the extra flux from the active Sgr A* (Kepler-10 b, CoRoT-24 b, 55 Cnc e, and EPIC 210894022 b). Such extreme wind temperatures are of course nonphysical because at temperatures $\sim 10^4$ K the gas would be fully ionized and in radiative equilibrium. In these conditions the gas heats up to $10^4$ K \citep{owen2012}. For any of the EPs in the sample whose calculated $T_{\textrm{XUV, p}}$ reaches temperatures $ > 10^4$, we assumed their $T_{\mathrm{min}}$ instead because this is in the thermostat temperature range for all EPs.

\subsection{Full-sky AGN sample}

\begin{figure}
\centering
\includegraphics[width=\hsize]{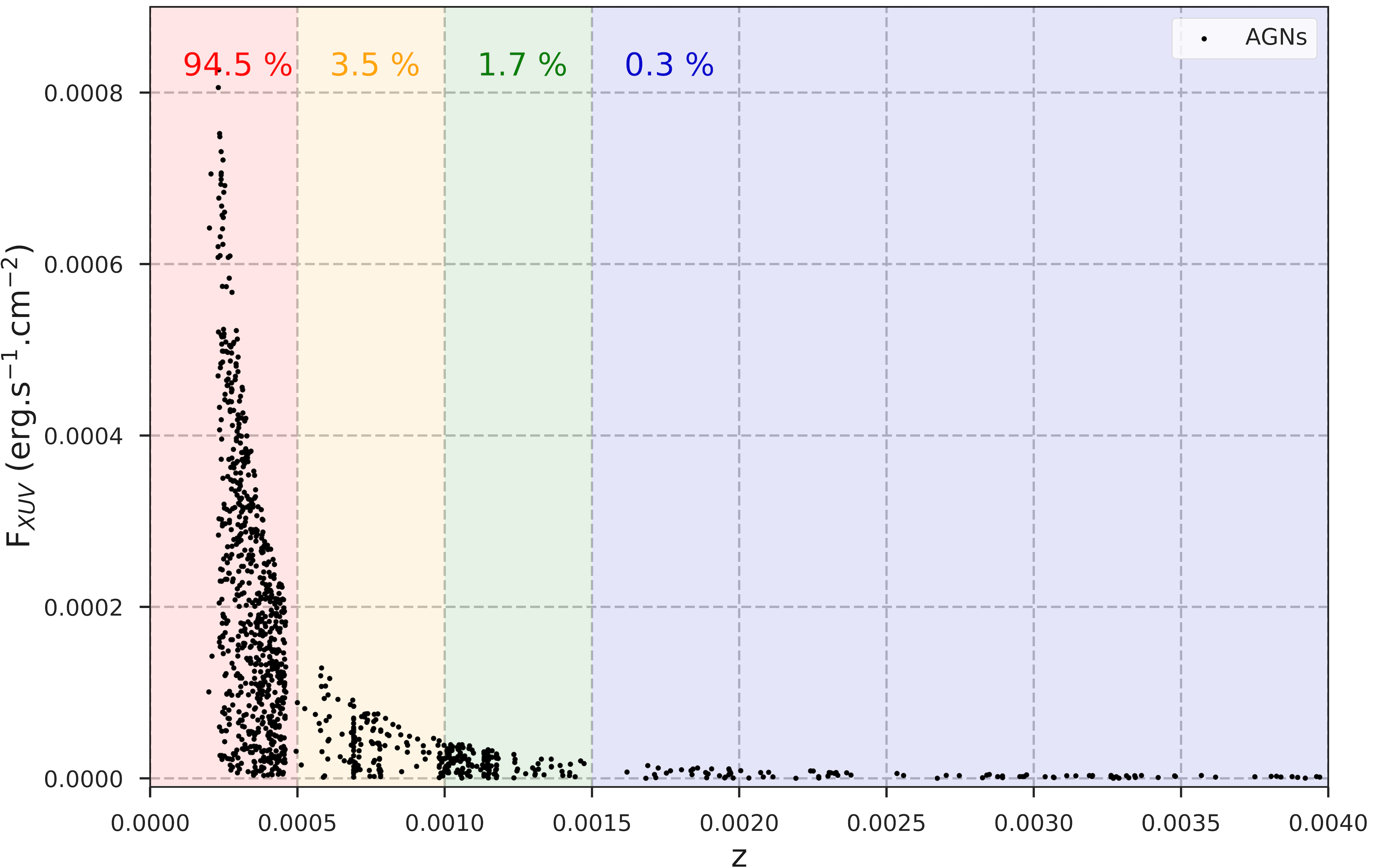}
\caption{Distribution of the generated $106,720$ AGN XUV fluxes with redshift. The shaded areas show the percentage of total flux enclosed in different ranges of $z$: $0 \leqslant z \leqslant 0.0005$ (red), $0.0005 \leqslant z \leqslant 0.0010$ (yellow), $0.0010 \leqslant z \leqslant 0.0015$ (green), and $z \geqslant 0.0015$ (blue). $99.7 \%$ of total flux reaches the MW from distances where $z \leqslant 0.0015$ and the majority of the AGNs, at large distances, account for only $0.3 \%$ of total flux.}
\label{AGN_flux}
\end{figure}

\begin{figure}
\centering
\includegraphics[width=\hsize]{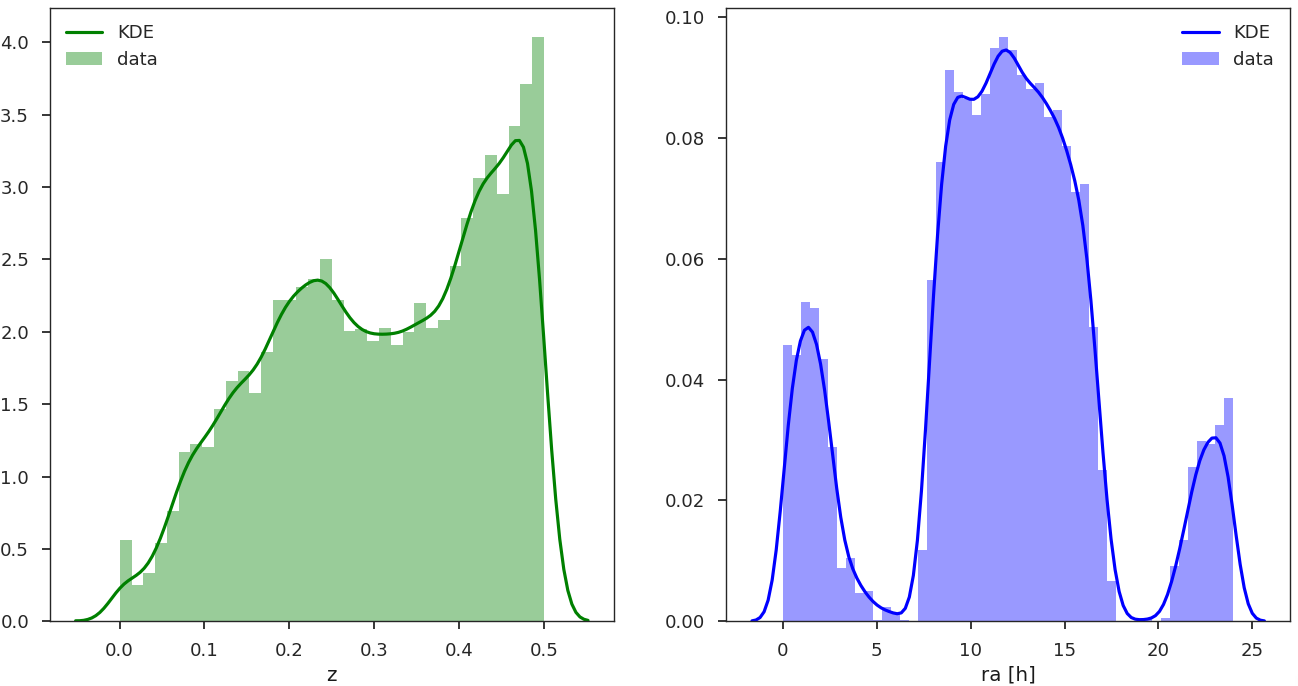}
\includegraphics[width=\hsize]{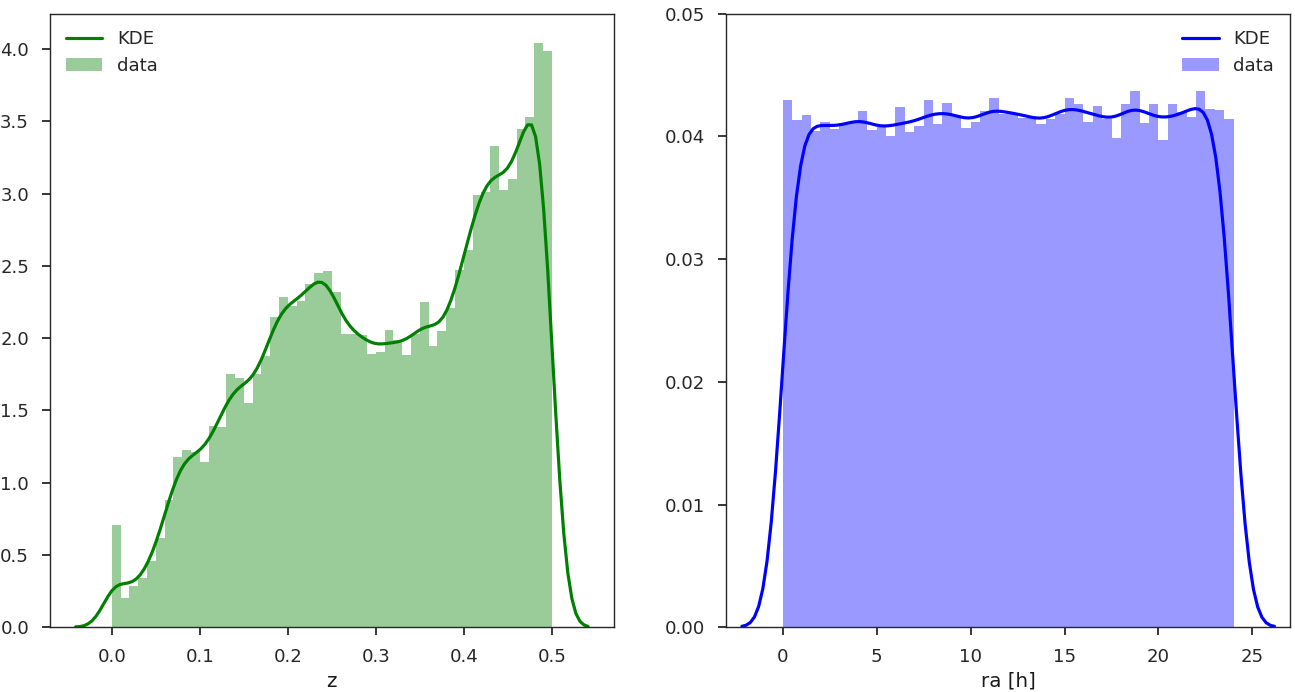}
\caption{ $z$ and $ra$ KDEs for the SDSS (top panel) and the generated (bottom panel) samples of AGNs. The distinct dips, due to the lack of data from these regions of the sky, can be seen in the ra KDE (top). Based on the SDSS sample distribution, we generated a new full-sky sample, with an assumed isotropic distribution (bottom).}
\label{KDEs}
\end{figure}

The list of AGNs was compiled from the SDSS observations, which cover only $\sim 15$ h of the sky \citep{paris2018, abolfathi2018}. Figure \ref{KDEs} (top) shows the kernel density estimates (KDEs) for the SDSS AGN distribution with $z$ (left) and $ra$ (right), where the distinct gaps due to the lack of data from these regions of the sky are observed in the KDE of $ra$. Upon counting the available SDSS AGNs, we have a rough estimate of $\sim 2,223$ AGNs/h. Because the AGN data are coarse, we generated a model of the AGN distribution covering the full sky ($106,720$), with an assumed isotropic distribution around the MW (see Figure \ref{KDEs} (top right)).

 We modeled the key features of the emission properties of AGNs using the standard AGN unified model. However, it should be noted that the exact physical processes that result in the observed emission of AGN are a highly dynamic topic of ongoing research. The standard physical model of an AGN starts with a central black hole that is growing through mass accretion. The accretion luminosity is given by $L_{\rm acc}=\eta \dot{M} c^2$, where $\eta$ is the mass-energy efficiency conversion (typically estimated to be 0.1) and $\dot{M}$ is the mass accretion rate.
A theoretical upper limit on $L_{\rm acc}$ can be calculated assuming accretion of fully ionized hydrogen onto a black hole of mass $M_{BH}$, balancing the force of gravity  to the radiation pressure from the accretion luminosity. This upper-limit is called the ``Eddington luminosity'' and has the value $\sim 1.3\times10^{38} \frac{M_{BH}}{M{\odot}}$ erg s$^{-1}$.

The standard AGN model is based on a black hole surrounded by an accretion disk. When this accretion disk is assumed to be optically thick (e.g., \cite{1973A&A....24..337S}) the resulting continuum spectrum is thermal with a
blackbody temperature of $\sim 2 \times 10 ^5$ K, which peaks in the UV
(10-400 nm) for a typical AGN (i.e., with $M_{BH}\sim 10^8 M{\odot}$).
Around the accretion disk is an optically thin corona
where the photons produced at the accretion disk pass through Compton upscattering to produce the power-law spectrum of high-energy X-rays of AGN.

The predominant presence of grains in the AGN torus that are larger than their interstellar counterpart has important astrophysical implications. In the AGN torus, larger grains mean a smaller amount of extinction (on a per unit dust mass basis) and a flatter
extinction curve in the  UV and optical wavelength range.
 \cite{2017ApJ...840...27S} have shown that the dust in AGN torus could exceed $\sim 0.4 \mu$m in radius, which is  substantially larger than that of the Galactic diffuse ISM  and
indicates a flat extinction curve for AGNs. For this reason, we chose to neglect its effects in the present work.

The result of our simple statistical analysis of the simulated sample of AGNs is given in Figure \ref{AGN_flux}, which shows the distribution of the $106,720$ AGN XUV fluxes with $z$. At $z \leqslant 0.0005$ (red shaded area in the plot), we find that $366$ objects (or $\sim 0.34 \%$) from the total sample contribute $94.5 \%$ of the total flux. In the range $0.0005 \leqslant z \leqslant 0.0010$ (yellow shaded) and $0.0010 \leqslant z \leqslant 0.0015$ (green shaded) we find $74$ (or $\sim 0.07 \%$) and $76$ (or $\sim 0.07 \%$) AGNs, respectively, which add up to $3.5 \%$ and $1.7 \%$ of the total flux. Thus, for the simulated sample, we find that $99.7 \%$ of the total flux comes from $z < 0.0015$. Conversely, the flux of the remaining $106,204$ AGNs at $z > 0.0015$ (blue shaded) adds up to only $0.3 \%$ of the total flux.

After accumulating all of the $106,720$ AGN fluxes, the $F_{\textrm{total}}/F_{\textrm{min}}$ ratio results in $\sim 0.8$. This is a striking result because it comes very close to $F_{\textrm{min}}$. We conclude that $\sim 500$ AGNs are missing in the range of $0 \leqslant z \leqslant 0.0015$ for $F_{\textrm{total}}$ to reach $\sim F_{\textrm{min}}$ and $\epsilon = 0.02$. Adding these missing AGNs to our sample, we end up with $107,220$ AGNs from which to calculate the mass loss.


\section{Results}


\subsection{Sgr A* and the SWEEPS planets}

\begin{figure}
\centering
\includegraphics[width=\hsize]{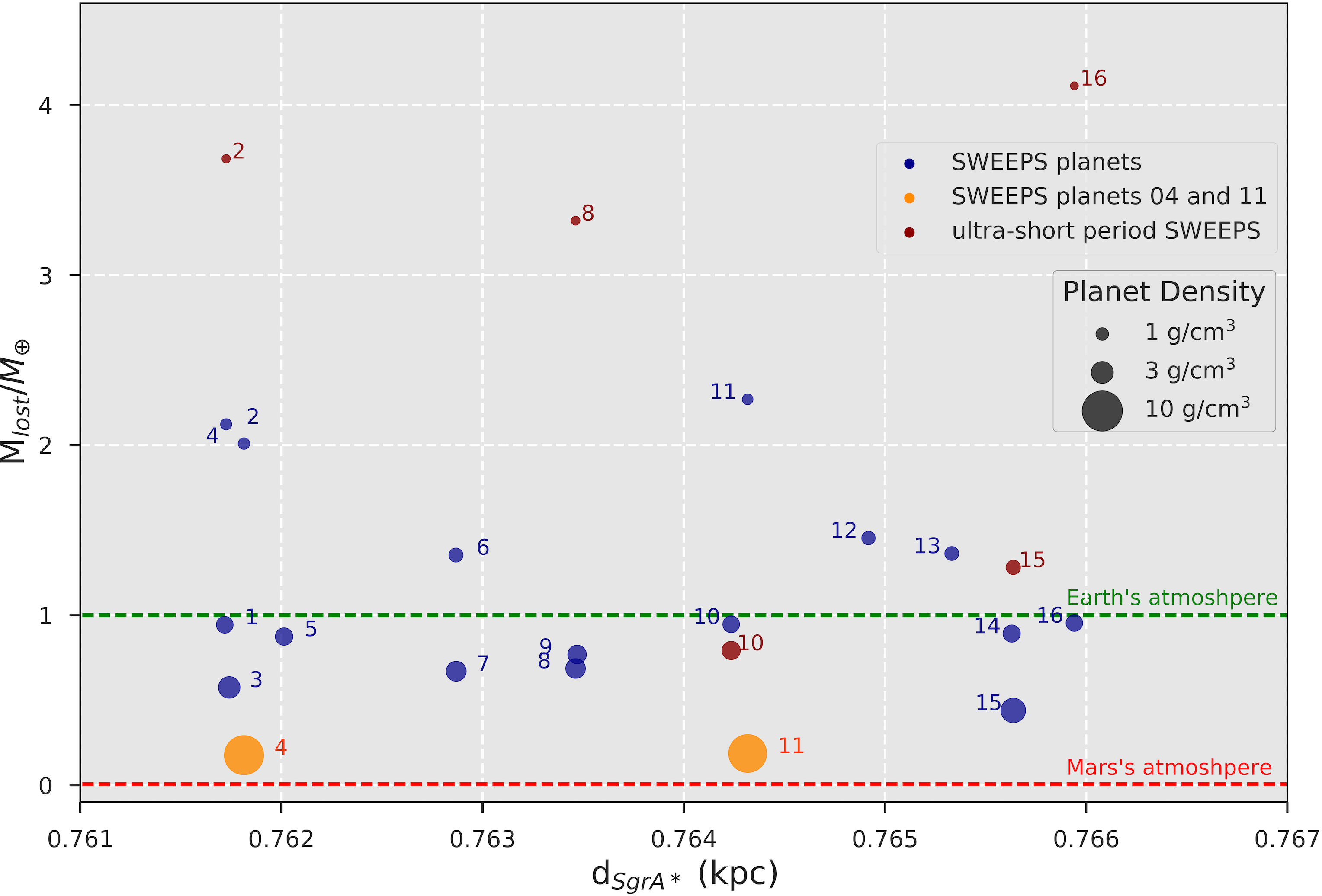}
\caption{Total mass loss of SWEEPS planet atmospheres in terms of Earth’s atmospheric mass as a function of distance from Sgr A*. The red points represent the ultra-short SWEEPS planets, and the yellow points show only the confirmed SWEEPS planets. The numbering of the points indicates each planet by name, hence some names are repeated in the plot. The green horizontal line indicates Earth's atmospheric mass, while the red horizontal line represents the mass of the Mars atmosphere.}
\label{Mlost_SWEEPS}
\end{figure}

\begin{table}
\caption{Total mass loss in the lifetimes of SWEEPS\ planets. Here the densities are determined from the masses estimated with the M-P anticorrelation (Section 2.1.1)}
\label{table:4.1}
\centering
\begin{tabular}{c c c c}
\hline
Name & $\rho$ (g cm$^{-3}$) & M$_{\textrm{lost}}$ (g) & M$_{\textrm{lost}}$ (\%)\\ \hline
1  & 1.77  & 4.73E+21  & 1.81E-07  \\
2  & 0.79  & 1.06E+22  & 3.68E-07  \\
3  & 2.90  & 2.88E+21  & 1.05E-07  \\
4  & 1.91  & 4.38E+21  & 3.02E-07  \\
5  & 1.23  & 6.78E+21  & 2.98E-07  \\
6  & 2.48  & 3.36E+21  & 1.71E-07  \\
7  & 2.42  & 3.44E+21  & 1.36E-07  \\
8  & 2.16  & 3.85E+21  & 1.32E-07  \\
9  & 1.75  & 4.74E+21  & 1.83E-07  \\
10  & 1.14  & 7.28E+21  & 2.34E-07  \\
11  & 1.21  & 6.83E+21  & 2.73E-07  \\
12  & 1.85  & 4.47E+21  & 2.23E-07  \\
13  & 3.76  & 2.20E+21  & 8.61E-08  \\
14  & 1.73  & 4.78E+21  & 2.40E-07  \\
15  & 0.83  & 1.01E+22  & 3.29E-07  \\
16  & 0.73  & 1.14E+22  & 3.97E-07  \\ \hline
\end{tabular}
\end{table}

\begin{table}
\caption{Total mass loss in the lifetimes of ultra-short SWEEPS planets. Here the densities are determined with the masses estimated with the P-M-R relation (Section 2.1.1)}
\label{table:4.2}
\centering
\begin{tabular}{c c c c}
\hline
Name & $\rho$ (g cm$^{-3}$) & M$_{\textrm{lost}}$ (g) & M$_{\textrm{lost}}$ (\%) \\ \hline
2  & 0.45  & 1.85E+22  & 1.11E-06  \\
8  & 0.50  & 1.66E+22  & 2.47E-06  \\
10  & 2.09  & 3.97E+21  & 6.95E-08  \\
15  & 1.29  & 6.42E+21  & 1.35E-07  \\
16  & 0.40  & 2.06E+22  & 1.31E-06  \\ \hline
\end{tabular}
\end{table}

\begin{table}
\caption{Total mass loss in the lifetimes of the SWEEPS 4 and 11 planets. Here the densities are determined with the real masses and radii of those planets \citep{sahu2007}.}
\label{table:4.3}
\centering
\begin{tabular}{c c c c}
\hline
Name & $\rho$ (g cm$^{-3}$) & M$_{\textrm{lost}}$ (g) & M$_{\textrm{lost}}$ (\%) \\ \hline
4  & 9.48  & 8.82E+20  & 1.22E-08  \\
11  & 8.92  & 9.32E+20  & 5.06E-09  \\ \hline
\end{tabular}
\end{table}

Figure \ref{Mlost_SWEEPS} shows the expected atmospheric mass loss in terms of Earth’s atmospheric mass in the lifetime of the SWEEPS planets as a function of distance from Sgr A*. This estimation includes the calculated mass loss that occurred during the active and quiescent stages of Sgr A*. In its AGN phase the $F_{\textrm{XUV}}$ reaching the SWEEPS planets is $\sim 4.7$ \textrm{erg s$^{-1}$ cm$^{-2}$}, with $F_{\textrm{XUV}} \setminus F_{\textrm{min}} \approx 47$ and the analogous efficiency $\epsilon = 0.1$. Since the end of its activity, the flux of Sgr A* arriving at the SWEEPS planets would be $\sim 1.5 \times 10^{-8}$ \textrm{erg s$^{-1}$ cm$^{-2}$}, resulting in $F_{\textrm{XUV}} \setminus F_{\textrm{min}} \approx 1.49 \times 10^{-8}$, whereby the XUV flux is much too low to result in atmospheric erosion. All of the mass loss therefore occurs during the AGN phase. The calculation was repeated for different mass estimations that are indicated in the plot. The blue points show the hot Jupiter M-P anticorrelation mass estimation we made for all of the SWEEPS planets. The red points represent the ultra-short hot-Jupiter mass estimation for the ultra-short SWEEPS planets (SWEEPS 2, 8, 10, 15, and 16), and the yellow points are the only confirmed SWEEPS planets (SWEEPS 4 and 11) with known masses and radii \citep{sahu2007}. The numbering of the points indicates each SWEEPS planet by name, hence some names are repeated in the plot. In the figure the green horizontal line represents Earth's atmospheric mass, $\sim 10^{21.7}$ \textrm{g} \citep{forbes2018}, and the red horizontal line represents the mass of atmosphere of Mars, $\sim 10^{19.4}$ \textrm{g} \citep{forbes2018}.

The density range of the blue data points lies between $0.73$ and $3.76$ \textrm{g cm$^{-3}$}, and the corresponding mass loss is $1.13 \times 10^{22} - 2.20 \times 10^{21}$ \textrm{g}. For the ultra-short SWEEPS 2, 8, and 16, the situation is more drastic, with a mass loss of about three to four times the Earth’s atmosphere ($1.85 \times 10^{22}$ \textrm{g}, $1.66 \times 10^{22}$ \textrm{g,} and $2.06 \time 10^{22}$ \textrm{g,} respectively), where their densities are $0.45$, $0.5$ and $0.40$ \textrm{g cm$^{-3}$}. The denser ultra-short SWEEPS 10 and 15 ($\rho_p = 2.09, 1.30$ \textrm{g cm$^{-3}$}) are expected to have lost $\sim 10$ times less of their atmospheres, $3.97 \times 10^{21}, 6.42 \times 10^{21}$ \textrm{g}. SWEEPS 4 and 11 ($\rho_p = 9.48, 8.92$ \textrm{g cm$^{-3}$}) experienced the lowest mass loss of $8.82 \times 10^{20}$ \textrm{g} and $9.32 \times 10^{20}$ \textrm{g,} respectively, which still exceeds the mass of the Mars atmosphere by nearly a factor of 4.

All the mass-loss values are more clearly represented in Tables \ref{table:4.1}-\ref{table:4.3}. The tables show the planetary density, the mass lost, and the percentage of mass lost in terms of their initial mass. We stress that the percentage of mass loss is calculated from the planet's initial bulk mass, not the atmospheric mass, for which we lack an estimate. However, in the case of Jovian planets, we can assume that a large portion of their bulk mass is contained in their atmospheres.

\subsection{Sgr A* and the Earth-like planets}

\subsubsection{H-only atmospheres}

\begin{figure}
\centering
\includegraphics[width=\hsize]{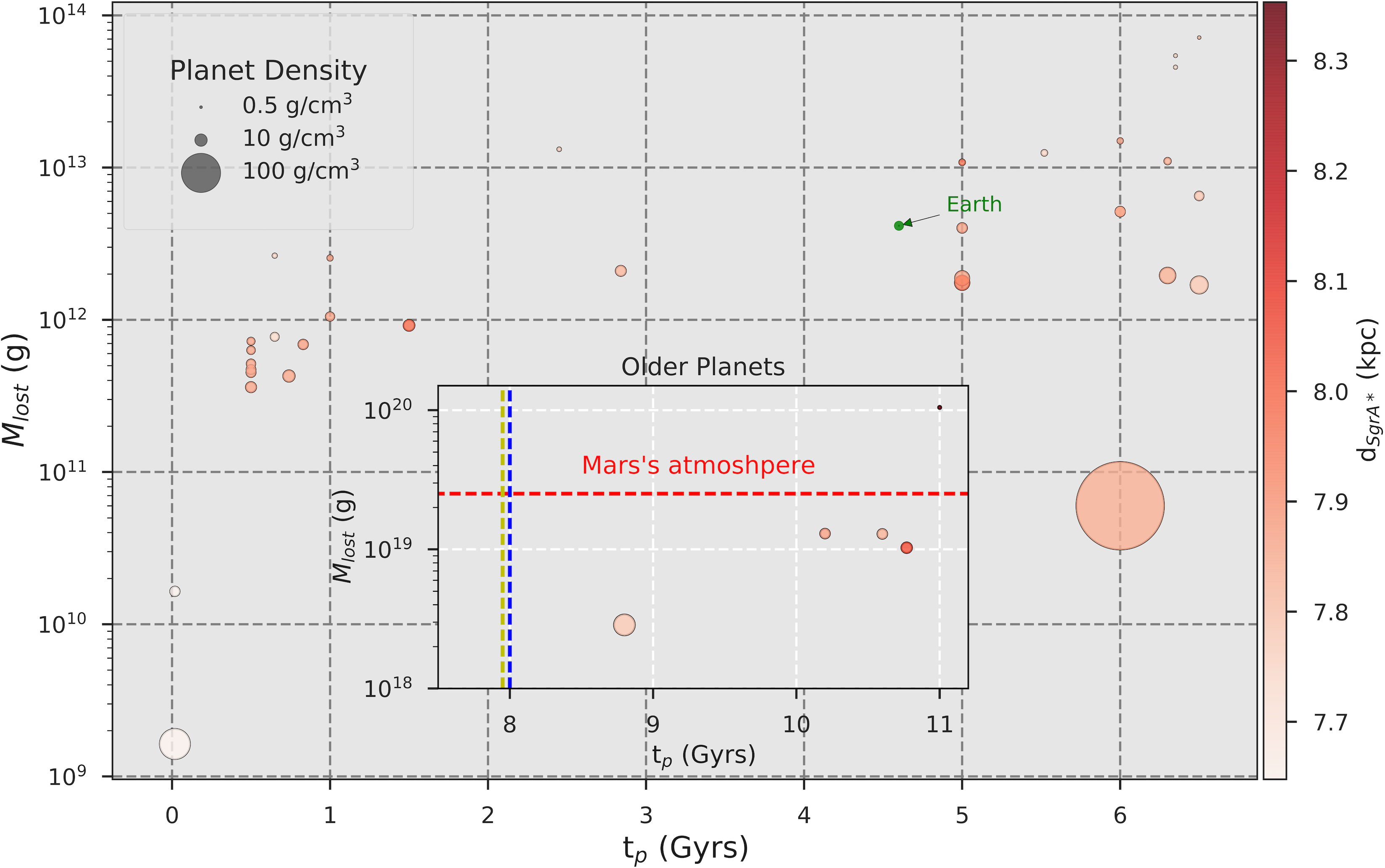}
\caption{Total mass loss experienced by the EPs in their lifetime as a function of age, with $\epsilon = 0.1$. main plot) the total mass lost by the younger EPs sample, inset subplot) total mass loss experienced by the older EPs. In the inset subplot, the blue dashed line marks the onset of the AGN phase of Sgr A* $8$ \textrm{Gyr} ago, the yellow dashed line represents the end of its activity $7.95$ \textrm{Gyr} ago, and the red dashed line signifies the mass of the Mars atmosphere. The color of the data points is indicative of the distance of the planet to Sgr A*.}
\label{Mlost_EPs}
\end{figure}

\begin{figure}
\centering
\includegraphics[width=\hsize]{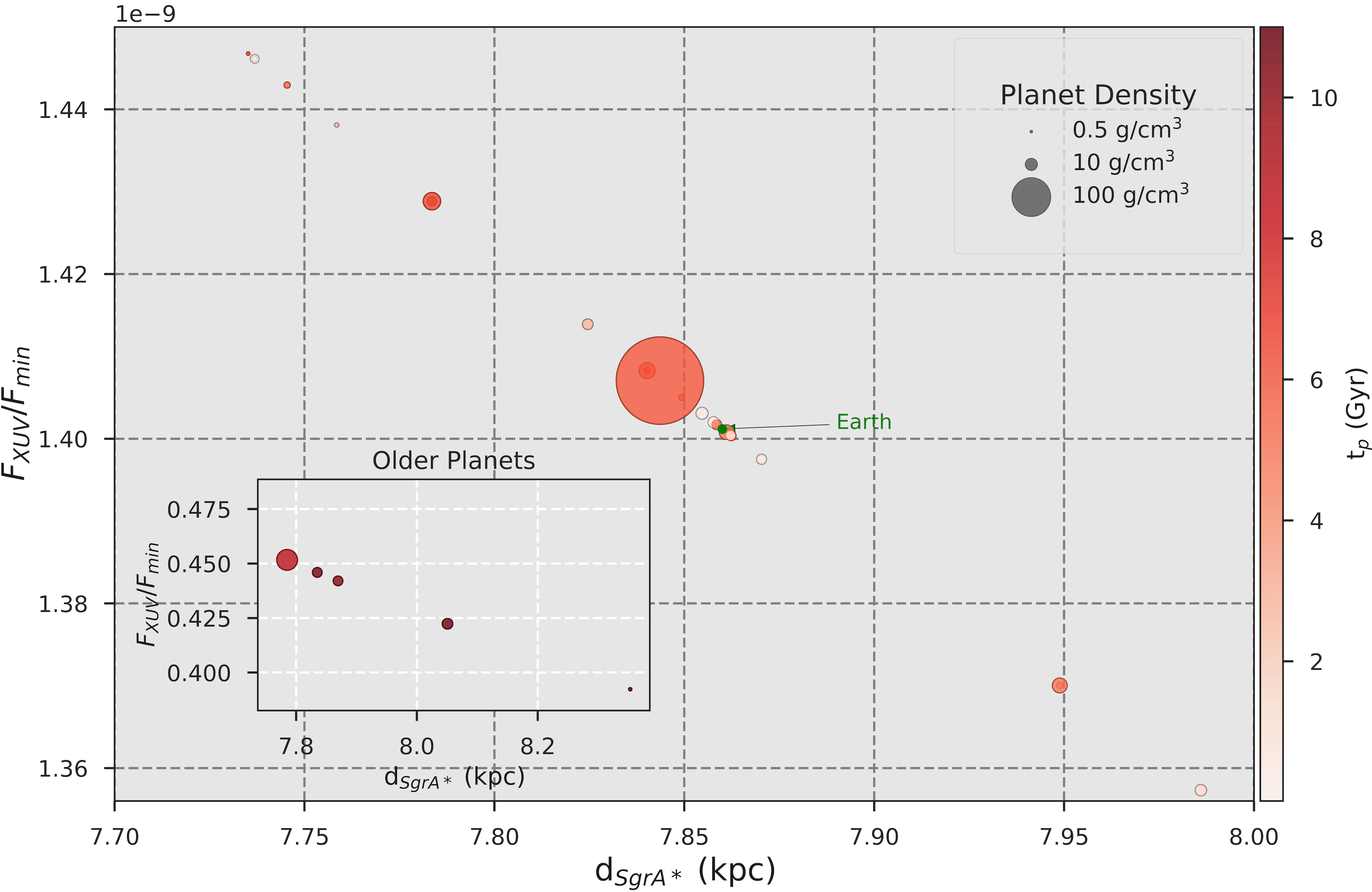}
\caption{Ratio of the flux arriving at the planet to the minimum flux required for atmospheric mass loss to occur, $F_{\textrm{XUV}}/F_{\textrm{min}}$, as a function of distance. The inset subplot shows results for the older planet sample during the AGN phase of Sgr A*, and the main plot presents the results for the entire EP sample during the quiescent phase of Sgr A*. Here the color of the points indicates the planet age.}
\label{Flux_ratio}
\end{figure}

\begin{table}
\caption{Total mass loss of the younger EPs ($ < 8$ Gyr) in their lifetimes. The mass loss is only due to the quiescent Sgr A*. For this result, $\epsilon = 0.1$.}
\label{table:4.4}
\centering
\begin{tabular}{l l c c}
\hline
Name   & Age (Gyrs) & M$_{\textrm{lost}}$ (g) & M$_{\textrm{lost}}$ (\%) \\ \hline
GJ 436 d   & 6       & 5.15E+12  & 3.19E-13   \\
GJ 436 c   & 6       & 5.12E+12  & 3.00E-13   \\
TRAPPIST-1 d & 0.5      & 4.71E+11  & 1.91E-14   \\
TRAPPIST-1 e & 0.5      & 5.15E+11  & 1.36E-14   \\
Kepler-70 c & 0.0184     & 1.64E+10  & 4.13E-16   \\
TRAPPIST-1 f & 0.5      & 7.23E+11  & 1.81E-14   \\
Kepler-100 c & 6.5      & 7.15E+13  & 1.39E-12   \\
TRAPPIST-1 b & 0.5      & 6.31E+11  & 1.23E-14   \\
Kepler-109 b & 6.35      & 5.44E+13  & 6.98E-13   \\
TRAPPIST-1 g & 0.5      & 4.51E+11  & 5.63E-15   \\
TRAPPIST-1 c & 0.5      & 3.61E+11  & 4.38E-15   \\
GJ 1132 b  & 5       & 4.01E+12  & 4.15E-14   \\
K2-3 c    & 1       & 2.55E+12  & 2.03E-14   \\
Kepler-68 c & 6.3      & 1.95E+12  & 1.50E-14   \\
Kepler-109 c & 6.35      & 4.56E+13  & 3.45E-13   \\
Kepler-37 b & 6       & 6.00E+10  & 3.62E-16   \\
Kepler-100 d & 6.5      & 6.51E+12  & 3.65E-14   \\
Kepler-289 c & 0.65      & 2.64E+12  & 1.07E-14   \\
Kepler-70 b & 0.0184     & 1.63E+09  & 6.15E-18   \\
CoRoT-7 b  & 1.5      & 9.19E+11  & 3.25E-15   \\
HD 3167 b  & 0.83      & 6.89E+11  & 2.30E-15   \\
Kepler-21 b & 2.84      & 2.10E+12  & 6.90E-15   \\
K2-141 b   & 0.74      & 4.27E+11  & 1.35E-15   \\
Kepler-92 c & 5.52      & 1.25E+13  & 3.68E-14   \\
K2-106 c   & 5       & 1.08E+13  & 3.17E-14   \\
Kepler-68 b & 6.3      & 1.10E+13  & 3.09E-14   \\
GJ 1214 b  & 6       & 1.50E+13  & 3.88E-14   \\
LHS 1140 b  & 5       & 1.88E+12  & 4.73E-15   \\
Kepler-289 b & 0.65      & 7.73E+11  & 1.77E-15   \\
Kepler-100 b & 6.5      & 1.69E+12  & 3.86E-15   \\
K2-3 b   & 1       & 1.05E+12  & 2.13E-15   \\
K2-106 b   & 5       & 1.75E+12  & 3.51E-15   \\
Kepler-88 b & 2.45      & 1.32E+13  & 2.54E-14   \\ \hline
\end{tabular}
\end{table}

\begin{table}
\caption{Total mass loss of the older EPs ($ > 8$ Gyr) in their lifetimes. The mass loss is due to the quiescent and active Sgr A*. For this result, $\epsilon = 0.1$, not true in reality, even when Sgr A* became an AGN.}
\label{table:4.5}
\centering
\begin{tabular}{l l c c}
\hline
Name     & Age (Gyr) & M$_{\textrm{lost}}$ (g) & M$_{\textrm{lost}}$ (\%) \\ \hline
Kepler-20 e   & 8.8      & 2.86E+18  & 1.55E-08   \\
Kepler-10 b   & 10.6      & 1.29E+19  & 6.49E-08   \\
CoRoT-24 b    & 11       & 1.05E+20  & 3.06E-07   \\
55 Cnc e     & 10.2      & 1.30E+19  & 2.68E-08   \\
EPIC 210894022 b & 10.77     & 1.03E+19  & 2.00E-08   \\ \hline
\end{tabular}
\end{table}

Figure \ref{Mlost_EPs} shows the total mass loss experienced by the EP sample in the EP lifetimes as a function of age, with $\epsilon = 0.1$. The main plot shows the total mass lost by the younger EP sample during the quiescent stage of Sgr A*. The inset subplot shows the total mass loss experienced by the older EP sample throughout the quiescent and active Sgr A* phases. In the inset subplot, the blue dashed line marks the onset of the AGN phase of Sgr A* $8$ \textrm{Gyr} ago, and the yellow dashed line represents the end of its activity $7.95$ \textrm{Gyr} ago. Again, the red dashed line signifies the mass of the Mars atmosphere. The color of the data points is indicative of the distance of the planet to Sgr A*. The mass loss shown for the younger planets (main plot) only includes the result for the quiescent Sgr A* flux because the AGN phase lasted only $50$ \textrm{Myr}, therefore the younger planets were born long after Sgr A* was active. When we assume $\epsilon = 0.,1$ the mass loss is as shown in Fig. \ref{Mlost_EPs}, spanning values from $10^{18}$ to $10^{20}$ \textrm{g} for the older planet sample, and the much less significant $10^9$ to $10^{14}$ \textrm{g} for the younger planets. Tables \ref{table:4.4} and \ref{table:4.5} show the mass lost by the younger and older EPs, respectively, in terms of their mass and percentage of bulk mass. While the distinguishing lines for the Jovian planets between the core and the atmosphere are not clealy defined, this is not the case for the terrestrial planets. In its lifetime, Earth would have lost $\sim 4 \times 10^{12}$ g or $\sim 7 \times 10^{-14}$ \% of its bulk mass. When we compare the mass lost to Earth's current atmospheric mass, the percentage value increases by $\sim 10^6$ and results in $\sim 8 \times 10^{-8}$ \% of Earths atmospheric mass lost in its lifetime so far as a result of the XUV radiation from the quiescent Sgr A*. Therefore, we expect the percentage of lost mass, in terms of planetary atmosphere, to be more significant than it appears in our tables.

When we follow the more careful approach and take $\epsilon = \epsilon(F_{\textrm{XUV}})$, we find that at no point in any of the EP lives was the $F_{\textrm{XUV}}$  strong enough to stir up a wind. Figure \ref{Flux_ratio} shows the ratio of the $F_{XUV}$ reaching the planets to the $F_{\textrm{min}} = 0.1$ \textrm{erg s$^{-1}$ cm$^{-2}$} required to bring $\epsilon$ up to $\sim 0.02$ and give rise to mass loss (Section 3.1.1) We observe that only the oldest planets in the sample were anywhere close to receiving $F_{\textrm{min}}$ during the active phase of Sgr A*  (inset subplot), with $F_{\textrm{XUV}}/F_{\textrm{min}}$ reaching $\sim 0.4$. However, they are still too far, at distances $\sim 7.8-8.3$ \textrm{kpc}, from the Galactic center to be in any way affected. In the quiescent phase of Sgr A*, we observe a range of $F_{\textrm{XUV}}/F_{\textrm{min}}$ between $\sim 10^{-9}$ for all the EPs. This ratio is again much too low to induce a wind.

\subsubsection{H/O atmospheres}
\begin{figure}
\centering
\includegraphics[width=\hsize]{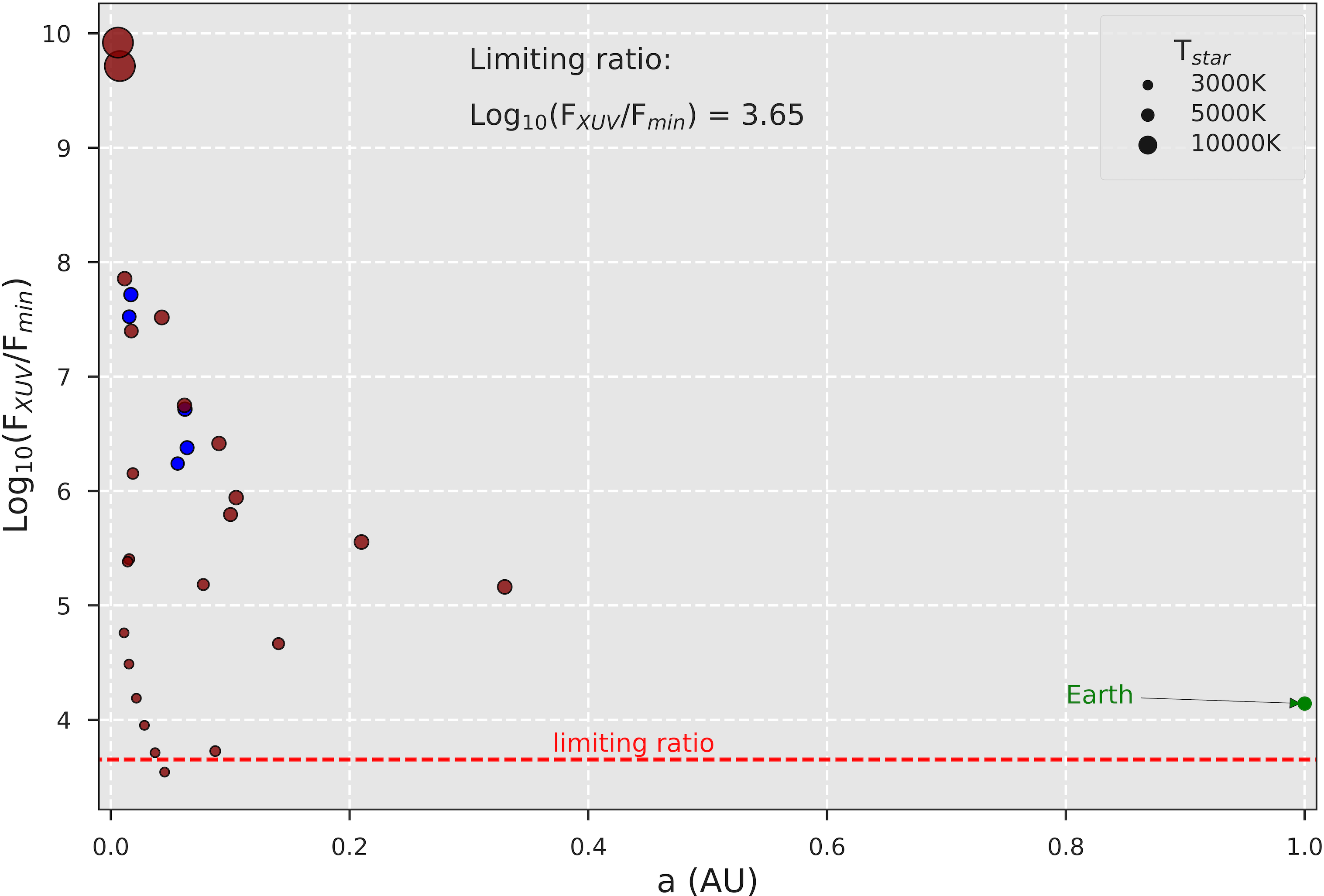}
\caption{Ratio of the sum of the stellar and Sgr A* XUV fluxes and $F_{\mathrm{min}} = 0.1$ erg s$^{-1}$ cm$^2$ plotted vs. planet semimajor axis. The red and blue points represent the younger and older planets, respectively. The limiting ratio of $\log_{10}(F_{\mathrm{XUV}}/F_{\mathrm{min}}) = 3.65$ (red dashed line) is the point at which the flow becomes increasingly radiation limited. The size of the point indicates the temperature of the EP host star.}
\label{Flux_ratios}
\end{figure}

\begin{table*}
\caption{First three columns: Calculated $T_{\mathrm{XUV}}$, $T_{\mathrm{min}}$ , and flux ratios of the younger EPs ($ < 8$ Gyr). Last column: Semimajor axes.}
\label{table:4.6}
\centering
\begin{tabular}{l l c c c}
\hline
Name       & $T_{\mathrm{XUV}}$ (K)   & $T_{\mathrm{min}}$ (K) & $\log_{10}\left(\frac{F_{\mathrm{XUV}}}{F_{\mathrm{min}}}\right)$ & a (AU)  \\ \hline
GJ 436 c   & 5.28E+06 & 2.23E+03 & 6.15 & 0.019 \\
TRAPPIST-1 d & 7.94E+03 & 2.76E+03 & 4.19 & 0.021 \\
TRAPPIST-1 e & 3.32E+03 & 3.58E+03 & 3.95 & 0.028 \\
Kepler-70 c & 4.27E+11 & 3.94E+03 & 9.71 & 0.008 \\
TRAPPIST-1 f & 2.08E+03 & 3.30E+03 & 3.71 & 0.037 \\
TRAPPIST-1 b & 4.71E+04 & 4.08E+03 & 4.76 & 0.011 \\
TRAPPIST-1 g & 6.12E+02 & 6.15E+03 & 3.54 & 0.045 \\
TRAPPIST-1 c & 9.25E+03 & 6.75E+03 & 4.49 & 0.015 \\
GJ 1132 b  & 1.87E+05 & 7.25E+03 & 5.40 & 0.015 \\
K2-3 c    & 4.41E+04 & 5.86E+03 & 4.67 & 0.141 \\
Kepler-68 c & 1.81E+06 & 1.22E+04 & 6.42 & 0.091 \\
Kepler-37 b & 8.56E+03 & 4.48E+04 & 5.79 & 0.100 \\
Kepler-289 c & 2.44E+05 & 7.97E+03 & 5.16 & 0.330 \\
Kepler-70  & 3.77E+10 & 3.02E+04 & 9.92 & 0.006 \\
CoRoT-7 b  & 6.26E+07 & 1.61E+04 & 7.40 & 0.017 \\
Kepler-21 b & 1.03E+08 & 1.60E+04 & 7.52 & 0.043 \\
K2-106 c   & 1.57E+06 & 1.20E+04 & 5.94 & 0.105 \\
Kepler-68 b & 1.74E+07 & 1.34E+04 & 6.75 & 0.062 \\
GJ 1214 b  & 2.64E+05 & 1.25E+04 & 5.38 & 0.014 \\
LHS 1140 b  & 2.79E+02 & 2.39E+04 & 3.73 & 0.088 \\
Kepler-289 b & 2.11E+05 & 1.76E+04 & 5.55 & 0.210 \\
K2-3 b    & 5.42E+04 & 1.96E+04 & 5.18 & 0.078 \\
K2-106 b   & 1.25E+08 & 2.83E+04 & 7.86 & 0.012 \\ \hline
\end{tabular}
\end{table*}

\begin{table*}
\caption{Same as Table \ref{table:4.6}, but for the older EPs ($ > 8$ Gyr).}
\label{table:4.7}
\centering
\begin{tabular}{l l c c c}
\hline
Name       & $T_{\mathrm{XUV}}$ (K)   & $T_{\mathrm{min}}$ (K) & $\log_{10}\left(\frac{F_{\mathrm{XUV}}}{F_{\mathrm{min}}}\right)$ & a (AU)  \\ \hline
Kepler-20 e   & 8.04E+05 &  1.84E+04   & 6.38   & 0.064 \\
Kepler-10 b   & 2.67E+08 &  1.17E+04  & 7.72   & 0.017 \\
CoRoT-24 b    & 1.33E+07 &  7.99E+03  & 6.24   & 0.056 \\
55 Cnc e     & 9.41E+07 &  2.10E+04 & 7.52   & 0.015 \\
EPIC 2(...) b & 5.36E+06 &   2.33E+04 & 6.72   & 0.062 \\ \hline
\end{tabular}
\end{table*}

In Figure \ref{Flux_ratios} we plot the ratio of the sum of the stellar and Sgr A* XUV fluxes and $F_{\mathrm{min}} = 0.1$ erg s$^{-1}$ cm$^2$ versus EP samimajor axis $a$.  The limiting ratio of $\log_{10}(F_{\mathrm{XUV}}/F_{\mathrm{min}}) =3.65$ is the point at which the flow becomes increasingly radiation limited and less energy limited, which requires a different approach. Because we include the stellar flux in this calculation and because of the small $a$ of the EPs, essentially all of the planets intercept fluxes that are too high for the wind to be treated as energy limited. Because there is no clear transition between the different flow regimes \citep{mclure2004, owen2015}, we continue to assume the energy-limited flow for the other three planets whose flux ratios are not too far above the limit: TRAPPIST-1 e, TRAPPIST-1 f, and LHS 1140 b ($\log_{10}(F_{\mathrm{XUV}}/F_{\mathrm{min}}) = 3.95$, $3.71$ and $3.73$). For the remaining planets in the sample, this is now a more inaccurate estimate because the radiation-limited treatment is beyond the scope of this work.

These four exoplanets also have calculated $T_{\mathrm{XUV}} < 10^4$ K, therefore the H/O mass loss is calculated with their $T_{\mathrm{XUV}}$ (see Table \ref{table:4.6}). In this case, the obtained mass loss in the lifetime of the planet is due to Sgr A* alone because the stellar flux was used solely to estimate the wind temperature. Keeping $\epsilon = 0.1$, the calculated $m_\mathrm{c} < m_\mathrm{O}$, meaning, no O is dragged along with the H flux, and for these planets, the mass loss is the same as shown in Figure \ref{Mlost_EPs} and Table \ref{table:4.4}. In the case of Earth, we obtain $\log_{10}(F_{\mathrm{XUV}}/F_{\mathrm{min}}) = 4.14$ and with $T_{\mathrm{XUV}} = 5035$ K, we also calculate $m_\mathrm{c} < m_\mathrm{O}$, therefore no loss of O atoms. For the remaining 24 EPs in the sample, the H/O mass loss was calculated using $T_{\mathrm{XUV}}$ when it is $\lesssim 10^4$ K or $T_{\mathrm{min}}$ otherwise (see Tables \ref{table:4.6} and \ref{table:4.7}). Similarly, with $\epsilon = 0.1$ and excluding the stellar flux, the calculated $m_\mathrm{c}$ is $< m_\mathrm{O}$, for all planets, therefore the mass loss is only in H and it remains the same as shown in Figure \ref{Mlost_EPs} and Tables \ref{table:4.4} and \ref{table:4.5}. The consequence of the planet losing H atoms only relates in fact to water loss because the O atoms left behind are unable to recombine again with the escaped H to form a water molecule.

\subsection{ Neighbor AGNs and the Milky Way Earth-like planets}

\begin{figure}
\centering
\includegraphics[width=\hsize]{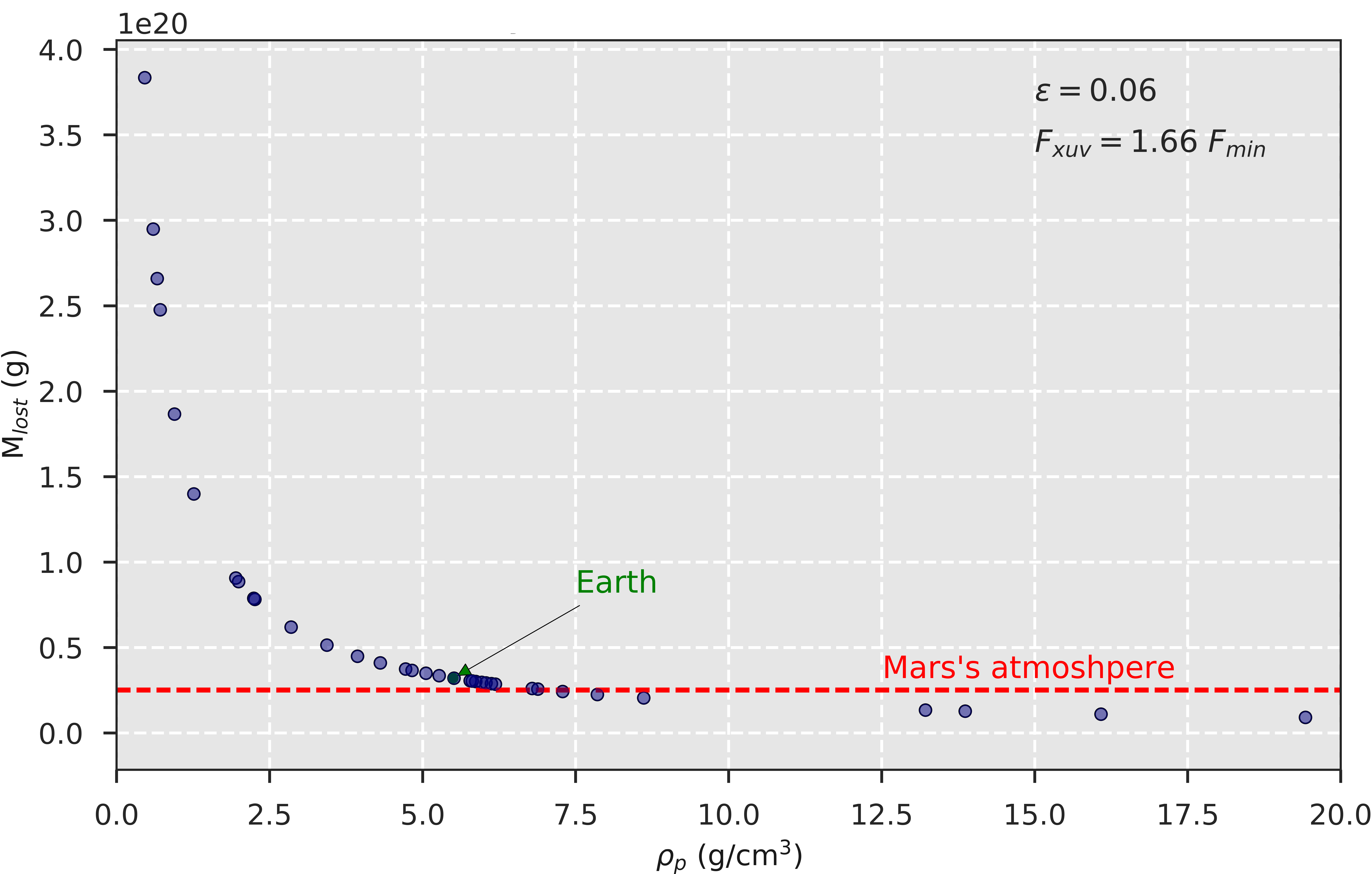}
\includegraphics[width=\hsize]{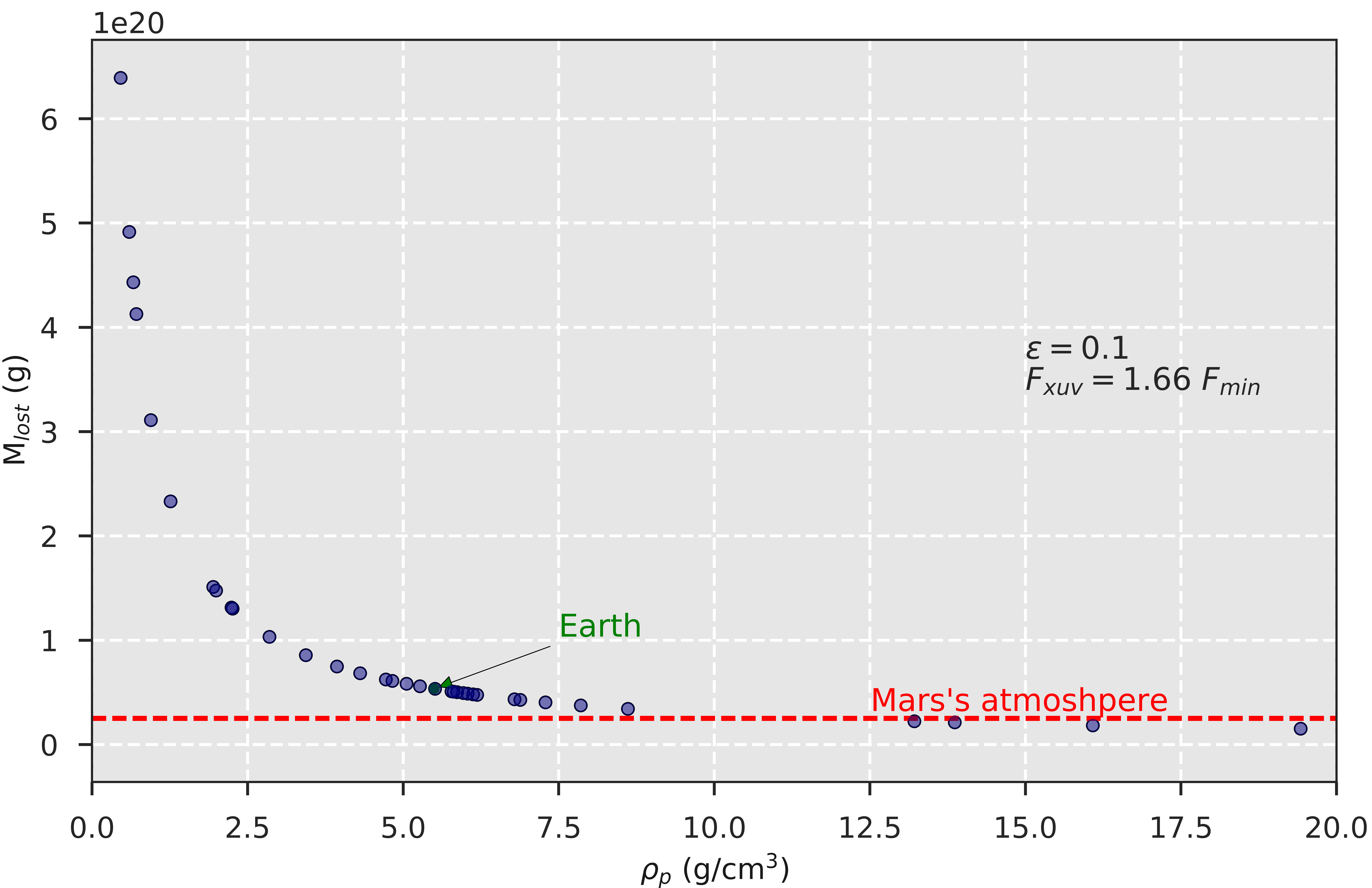}
\caption{ (Top) The total mass lost vs EP density over a period of 50 \textrm{Myrs} of irradiation by $107,220$ AGNs, up to $z = 0.5$. The $F_{\mathrm{XUV}}$ arriving at the MW amounts to $\sim 166 \%$ of the $F_{min}$, with the corresponding efficiency of $\epsilon = 0.06$. The densest planets in the sample are not shown in the graph for clarity. The red, horizontal line indicates the mass of atmosphere of Mars. (Bottom) same as top, but with $\epsilon = 0.1$.}
\label{Mlost_AGNs}
\end{figure}

For a total sample of $107,220$ AGNs we obtain $F_{\textrm{total}} = 1.66$ $F_{\textrm{min}}$, with the corresponding efficiency of $\epsilon = 0.06$ \citep{bolmont2016}. With these values of $F_{\textrm{total}}$ and $\epsilon,$ we estimated the total atmospheric mass loss over the period of $50$ \textrm{Myr} for the EPs sample in the MW. Figure \ref{Mlost_AGNs} shows our results of the total mass loss as a function of planet density. In the figure (top) we see that over the period of $50$ \textrm{Myr} of XUV irradiation by $107,220$ AGNs, Earth is expected to lose $3.24 \times 10^{19}$ \textrm{g} of mass. This amounts to $\sim 8.47 \times 10^{-7} \%$ of its bulk mass or $\sim 0.6$ \% of its current atmospheric mass. The lowest density planets ($\rho_p \leq 1.5$ \textrm{g cm$^{-3}$}) show much more drastic erosion, spanning $1.3 \times 10^{20} - 3.7 \times 10^{20}$ \textrm{g}, whereas the densest planets ($\rho_\textrm{p} \geq 12$ \textrm{g cm$^{-3}$}) lose only about $3.6 \times 10^{17} - 1.4 \times 10^{19}$ \textrm{g} of mass.

If we increase the efficiency coefficient to $\epsilon = 0.1$, we observe higher mass loss by a factor of $\sim 1.6$, which we show in Figure \ref{Mlost_AGNs} (bottom). With this exaggerated $\epsilon,$ the mass lost by Earth is $5.45 \times 10^{19}$ \textrm{g}, which amounts to $\sim 9.13 \times 10^{-7}$ \% of its bulk mass or $\sim 1$ \% of its current atmospheric mass. The lowest density planets ($\rho_\textrm{p} \leq 1.5$ \textrm{g cm$^{-3}$}) experience mass loss of $2.4\times 10^{20} - 6.4\times 10^{20}$ \textrm{g}, and the densest planets ($\rho_\textrm{p} \geq 12$ \textrm{g cm$^{-3}$}) lose $6 \times 10^{17} - 2.3 \times 10^{19}$ \textrm{g} of bulk mass.

The effect of SMBH  radiation  on the known systems in the Galactic solar neighborhood is difficult to consider because we lack  data:
Proxima Centauri b, in orbit of Proxima Centauri (d$_{Earth} = 1.295$ pc), with a mass of $m\sin{(i)} = 1.3$ M$_{\oplus}$, is the closest potentially habitable planet to the solar system. To date, the radius of this planet still remains to be established. Without this crucial information, we are unable to estimate the potential mass it may have lost through XUV flux of Sgr A* and the neighboring AGN. However, assuming R$_{\oplus}$ for Proxima Centauri b, we deduce for it an approximate mass loss of $0.8 M_{lost,\oplus}$. Because the age of  Proxima Centauri is unknown, we adopt for it the age of the solar system ($4.6$ Gyr) assuming it was born from
the same molecular cloud. Given its distance from the Galactic center, it is unlikely that Proxima Centauri b was affected by Sgr A* in its lifetime. In terms of the XUV radiation from the surrounding AGNs, its expected bulk mass loss amounts to $\sim 3.46 \times 10^{-7} \%$  ($\epsilon = 0.06$) and $5.77 \times 10^{-7} \%$  ($\epsilon = 0.1$.

\section{Discussion}
Before proceeding with the discussion, we briefly summarize the assumptions we used to analyze the radiation effects of Sgr A* and $z < 0.5$ AGNs:

\begin{itemize}
\item no photon losses, that is, $\tau = 0$ everywhere,

\item $L_{\textrm{Edd}}$ and $\eta_\textrm{X} = 0.7$ for all AGNs,

\item accretion period of 50 Myr for all AGNs,

\item random mass from the distribution of $ M_{\textrm{BH}} = 10^{7-9} M_{\odot}$ for all AGNs,

\item a spatial distribution of $z$ of the AGNs as obtained from the SDSS DR14,

\item H-only atmospheres for all planets, with the single exception of H/O atmospheres for the EPs,

\item the same distance to all extragalactic AGNs for all MW planets.

\end{itemize}

\subsection{Sgr A effects}
To date, no terrestrial exoplanets are known to exist within the galactic bulge.
Instead, only a handful of hot Jupiters is detected in close proximity to the galactic center. Our study of the SWEEPS planets shows that the maximum mass loss experienced by the low-density hot Jupiters at distances $d < 0.7$ \textrm{kpc} could be as high as $4 M_{\textrm{Mars}}$ by the end of the Sgr A* activity (Figure \ref{Mlost_SWEEPS}). Conversely, by our estimates, none of the EPs would have experienced any atmospheric erosion, with distances $ > 7.8$ \textrm{kpc}, regardless of their bulk density. Although all of the EPs in our sample are at distances greater than would imply any atmospheric damage to their atmospheres through the activity of Sgr A*, we cannot exclude the possibility that some of the stars of these planets migrated within the Galaxy \citep{loebman2016}, in which case the EP stars could have been much closer to Sgr A* $\sim 8$ \textrm{Gyr} ago, where the EPs could have experienced some atmospheric erosion after all. On the other hand, \citet{prieto2004s} have explored this scenario for the Sun and concluded that it most likely formed at its current galactocentric distance.

Previous studies, for example, \citet{forbes2018}, concluded that terrestrial planets at distances greater than $\sim 0.1$ \textrm{kpc} from Sgr A* are not likely to be affected with significant mass loss during the active phase of Sgr A*. \citet{balbi2017} estimated the cap at $\sim 1$ \textrm{kpc}, with mass loss comparable to that of Earth's current atmosphere at distances $d \leq 0.5$ \textrm{kpc}. However, these studies did not use real exoplanet data in their analysis, but treated the problem statistically and considered only terrestrial planets with densities comparable to that of Earth, $\rho_{\oplus} = 5.5$ g cm$^{-3}$.

The assumption of pure H atmospheres for all our planets is not accurate, in particular concerning the EPs. This assumption is more relevant for the gas giant SWEEPS planets. The energy required to lift H out of the planetary gravitational potential is much lower and easier to attain than heavier atmospheric constituents. Considering a typical terrestrial atmosphere, we will find many more heavy elements and molecules (e.g., H$_2$O, CO$_2$ , and CH$_4$). Our investigation included the consideration of an H/O atmosphere, that is, the presence of water, being subjected to XUV flux from the active Sgr A*. We find that the atmospheric heating due to the combined SMBH and stellar fluxes would not in fact result in the O drag with the escaping H. Our result implies that as a consequence, after the H$_2$O molecules have been dissociated, the H escapes, leaving O behind, with no chance of recombining into H$_2$O again, which leads to water loss \citep{bolmont2016}. In addition to these simplifications, the resulting mass loss in H would give us an overestimated value because we did not account for energy loss (reflected in $\epsilon$) resulting from the photodissociation of water.

There are claims that some or even significant atmospheric erosion could be beneficial for the emergence of life, for example, in the case of mini-Neptunes \citep{luger2015}. Furthermore, there are indications that an atmosphere has the potential of being replenished as the result of ongoing geological events such as outgassing and volcanism, on whose account \citet{balbi2017} suggested incorporating an additional factor into Equation \ref{eq:mass_lost_total}. \citet{marty2012} used these arguments to explain how water trapped in the mantle can replenish the surface water content. On planets with no substantial atmospheres, the development or survival of complex life could be in significant danger, even at distances $\sim 1$-$10$ \textrm{kpc} from the active Sgr A*, in terms of lethal doses of ionizing radiation infringed upon living organisms \citep{balbi2017}.

\subsection{AGN effects}
Considering the cumulative effect of all the surrounding AGNs ($z < 0.5$) on atmospheric erosion of the MW EPs, we see a different picture. We stress yet again that the inclusion of the external AGNs as possible sources capable of causing atmospheric erosion has so far never been acknowledged. The sample of $107,220$ AGNs was obtained even though observational data of the SDSS DR 14 were lacking severely, which means that many more (i.e., $\sim 500$) AGNs could exist in the nearby volume of space. It is also quite unlikely that our generated sample fully resembles the AGN distribution in the neighboring universe. With all this considered, we are surely looking at a large margin of error in our mass-loss estimates. Upon adding the $\sim 500$ AGNs with redshifts $\leq 0.0015$ to our sample, the total XUV flux from $107,220$ AGNs increased to $1.66 F_{\textrm{min}}$. This is a striking result because this flux is enough to give rise to a hydrodynamic wind, and so we expect that the MW planets could have been subject to enough XUV radiation to induce atmospheric erosion in their past, but it could also be an ongoing process in the present.

We assumed $\tau = 0$, so that we can expect the influence of the surrounding AGNs to be of lesser significance than what our results show. So far, \citet{balbi2017} have studied the problem of galactic absorption by assuming optical depths of $\tau = 1$ in the Galactic plane and $\tau = 0$ at greater angles to it. Future studies will benefit from similar considerations, ensuring better constraints on the expected atmospheric mass loss of the MW planets due to the proximal AGNs. From the analysis of \citet{mclure2004} it emerges for now that most of the AGNs show broad lines in their spectra, implying a more face-on inclination and therefore $\tau \ll 1$. Studying them also sheds more light onto our inaccurate assumption of $L_{\textrm{Edd}}$ for all AGNs because most AGNs at $z \leqslant 0.5$ right now accrete at rates lower than this \citep{mclure2004}. Future observations with the Large Synoptic Survey Telescope (LSST) or the European Extremely Large Telescope (E-ELT) will provide more detailed information on the nature and characteristics of AGNs and help greatly in more detailed atmospheric erosion estimations.

\subsection{Effects of galaxy mergers}
We would like to reflect on possible causes contributing to the atmospheric erosion because this clarifies the evolution of planetary atmospheres on cosmological timescales.
Galaxy mergers are seen in the local and distant universes. They may affect the formation of SMBH and trigger AGN activity, as both observations and theory have shown \citep[e.g.,][]{bundy2008, hopkins2008}. These events are relevant for the future of the MW because it is bound to eventually collide with the Andromeda galaxy \citep{cox2008, sohn2012m31}, although there is no evidence of the MW undergoing a major merger in the last $10$ Gyr \citep{stewart2008}. During such mergers the SMBHs may be most active, resulting in the enhancement of the XUV emissions. \citet{van2012m31} predicted that after the merger, stars at similar galactocentric distance as the Sun will have significantly larger orbits than $8$ kpc, which places them at a safe distance from the XUV radiation emitted from the central AGN. Additionally, we have now found that the neighboring AGNs should be included in assessing the overall damage that planets might experience in their lifetimes. The problem becomes even more complex as we consider the ever-changing scale of the universe and the growing distances between galaxies. We have to keep in mind that at the time Sgr A* went though its active phase, other AGNs would exist in the nearby universe, adding their XUV flux to that of our galactic nucleus. This would have made the total XUV flux far more effective at the time.

\section{Conclusions}
We presented a comparative analysis of the influence of Sgr A* and nearby AGNs ($z<0.5$) on the atmospheric mass loss of Earth-like and SWEEPS planets. For the first time we coupled exoplanet and AGN data with theoretical models, giving the first insight into the threatening effects AGNs could pose to planetary atmospheres at present and in the past. For this goal, we used the simple analytical estimations of atmospheric mass loss in the energy-limited regime. A full numerical treatment of radiative transfer, conduction, and photochemical modeling of the atmosphere incorporated by programs such as ZEUS or the PLUTO-CLOUDY interface will be addressed in future work. We draw the following conclusions from our study:

\begin{itemize}
\item Considering the close proximity of the 16 SWEEPS planets to Sgr A* ($\sim 0.761-0.766$ kpc), their age ($\sim 10$ Gyr), and the onset of activity of the MW core $\sim 8$ Gyr ago, we determined the upper limit for the mass loss in their lifetime as the result of XUV flux from the active and quiescent Sgr A*. During the active phase, the XUV flux arriving at these planets is $\sim 4.7$ \textrm{erg s$^{-1}$ cm$^{-2}$}, with $F_{\textrm{XUV}}/F_{\textrm{min}} \approx 47$ and $\epsilon = 0.1$. Our results show that the low-density hot Jupiters at distances $d < 0.7$ experienced the maximum mass loss, possibly as high as 4 $M_{Mars}$ by the end of the Sgr A* activity. In terms of the bulk atmospheric mass of these planets, this result is quite insignificant, however, which is clearly depicted in Tables \ref{table:4.1}-\ref{table:4.3}. We observe no mass loss related to Sgr A* in its quiescent phase because we also showed that $F_{\textrm{XUV}} \sim 1.5 \times 10^{-8}$ \textrm{erg s$^{-1}$ cm$^{-2}$}, giving $F_{\textrm{XUV}}/F_{\textrm{min}} \approx 1.49 \times 10^{-8}$ for all SWEEPS planets. In these circumstances, the XUV flux is much too low to result in atmospheric erosion.

\item In the case of the EPs with H-only atmospheres, we found that at no point in their lifetimes were any of the planets susceptible to atmospheric mass loss caused by the XUV radiation from Sgr A*. For the younger planet sample ($< 8$ Gyr), which includes Earth and the five potentially habitable exoplanets, it is more obvious as $F_{xuv}/F_{min} \approx 1.4 \times 10^{-9}$, which is much too low to give rise to a hydrodynamical wind. For the older planets ($> 8$ \textrm{Gyr}), the situation is very different, with $F_{\textrm{XUV}}/F_{\textrm{min}} \approx 0.4$ during the active phase of Sgr A*, this ratio still being too low to induce atmospheric erosion because all the planets in the sample are at safe distances to our Galactic center, with $d_{\textrm{Sgr A*}} \sim 7.8 - 8.3$ \textrm{kpc}.

\item We found that at its current luminosity, the XUV flux from Sgr A*, even with the assumption of $\eta_\mathrm{X} = 0.7$, is unable to induce a hydrodynamic wind in any of the planets (EPs and SWEEPS) in our sample. For this reason, the atmospheric erosion due to an inactive Sgr A* does not need to be regarded as a hazard for life, even at very short distances.

\item Considering H/O atmospheres of the 28 EPs, our results indicate that after the dissociation of water molecules, the Sgr A* XUV flux would result in H escape, leading to water molecule loss. With the wind temperatures due to the host star and Sgr A* calculated for the four exoplanets with habitable zones (LHS 1140 b, TRAPPIST-1 e, f, and g), we found that when $\epsilon = 0.1$ and $\eta_{\mathrm{X, SgrA*}} = 0.7$, the quiescent Sgr A* XUV flux is causing H mass loss from their atmospheres, but not O drag. For the remainder of the EP sample, the combined stellar and Sgr A* XUV flux is too high for the wind to be considered energy limited. However, with our energy-limited estimates, none of the planets loses any O. In general, combined with stellar flux, the XUV flux Sgr A* could be contributing to atmospheric erosion.

\item What we see from the study of the neighboring AGNs ($z < 0.5$) is the result of the combined $F_{\textrm{XUV}}$ of the generated sample of $107,220$ AGNs. Their $F_{\mathrm{total}} = 0.165$ erg s$^{-1} $cm$^{-2}$, and $\epsilon = 0.06$. In such circumstances, Earth is expected to lose $\sim 8.47 \times 10^{-7} \%$ of its bulk mass or $\sim 0.6 \%$  of its current atmospheric mass. With $\epsilon = 0.1$, the mass lost by Earth results in $\sim 9.13 \times 10^{-7} \%$  of its bulk mass or $\sim 1  \%$ of its current atmospheric mass. We lack knowledge about the atmospheric mass of the exoplanets in our EP sample, but we can conservatively conclude that a similar result is expected for the remaining EPs. Our result is of special interest for the EPs orbiting within the habitable zone of their stars (TRAPPIST-1 d, e, f, g, and LHS 1140 b). The external AGN activity might be a factor worth considering in terms of its influence on shaping of the atmospheres and influencing planetary habitability with time.
  Here, our underlying assumption was that $\tau = 0$, in conjunction with all of the nuclei becoming AGNs at the same time. We therefore conclude that the atmospheric mass loss, as seen in Figure \ref{Mlost_AGNs}, is the upper limit for the EP sample for a timescale of 50 Myr because they were already formed when the AGN flux began. Because galaxy mergers can affect AGN activity and the MW is on an eventual collision route with Andromeda galaxy, we may expect that the activity in Sgr A* increases when the collision occurs. In the past time of the SgrA* active phase, other AGNs would also have existed, adding their XUV flux to that of our galactic nucleus. This would have made the total XUV flux far more effective.

\end{itemize}

The results obtained in this work are forerunners of those that we can expect to obtain with the success of future missions, like the James Webb Space Telescope (JWST), the Atmospheric Remote-sensing Infrared Exoplanet Large-survey (ARIEL), and the E-ELT. These missions will be able to provide us with detailed analysis of exoplanet atmospheres, which in turn could translate into better estimates of erosion in terms of atmospheric mass rather than bulk mass of the planets. In terms of providing a better map of the AGNs and galaxies around the MW, the upcoming data from the E-ELT will also be very beneficial. With this information, a clearer assessment of the external factors that influence the time evolution of MW planets will be possible, in particular, with better treatment of photon absorption within the galaxies.

Our analysis based on real exoplanet and AGN data confirms that to determine the evolution of exoplanet atmospheres, it is essential not only to take into account the impact of radiation from their host star, but also from the host galaxy and external AGN emission, with possible consequences for the origin of life and its development.

\begin{acknowledgements}
This work is supported by projects (176001) Astrophysical Spectroscopy of Extragalactic Objects of Ministry of Education, Science and Technological Development of Serbia, and by the Erasmus Mundus Master Program, AstroMundus.
\end{acknowledgements}

\bibliographystyle{aa} 
\bibliography{bibliography} 

\begin{appendix}

\section{Mass estimation of the SWEEPS planets}

\subsection{Hot Jupiters}

\begin{figure}
\centering
\includegraphics[width=\hsize]{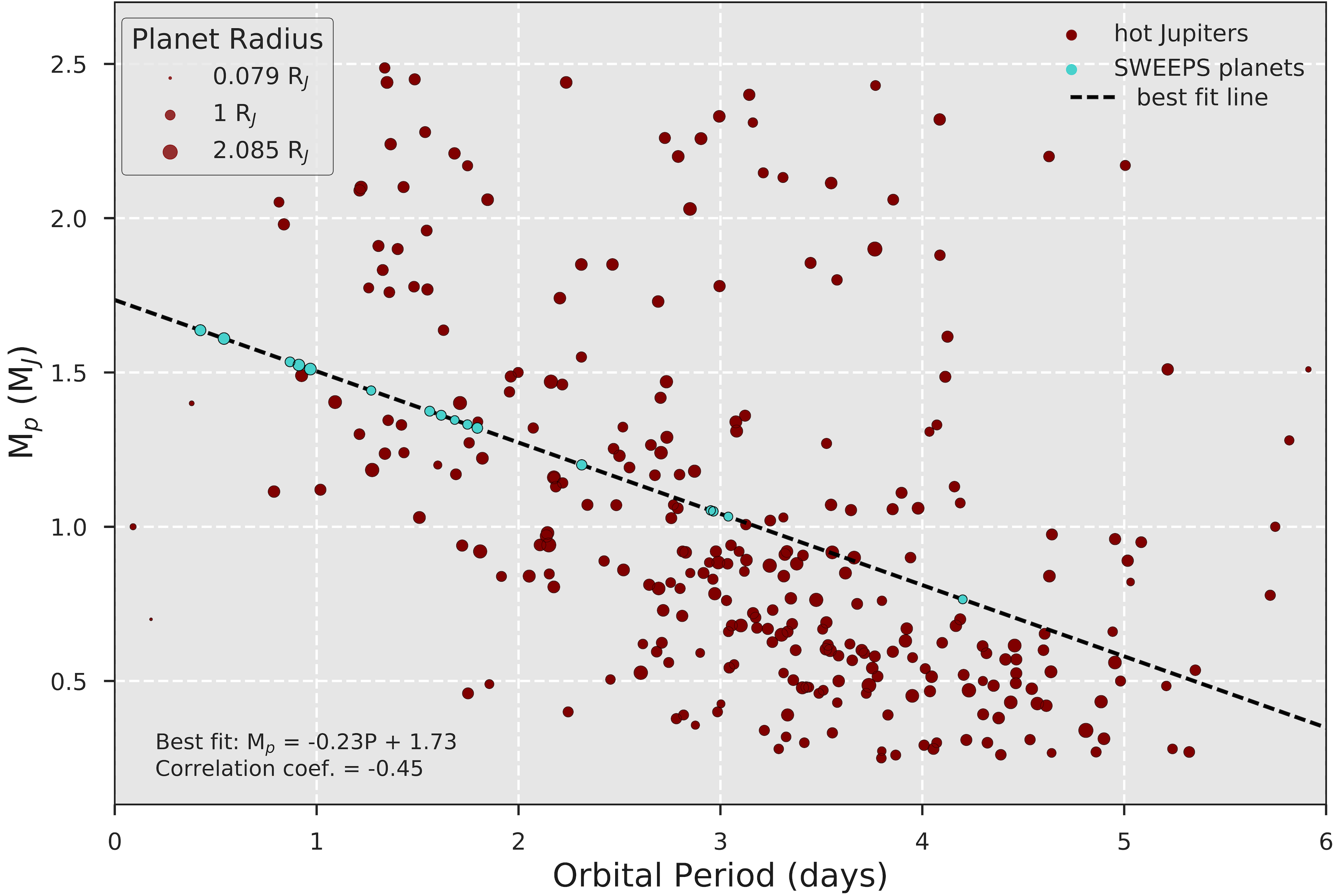}
\caption{M-P relation for the hot Jupiters used to estimate the masses of the SWEEPS planets, which is necessary for the mass-loss estimation. The red points represent all the hot Jupiters ($P < 10$ \textrm{days}) discovered to date (312 exoplanets) and available in the EPE database. The black dashed line is the best-fit line, $M_\textrm{p} = -0.23 P + 1.73$, for this distribution. The masses of the 16 SWEEPS planets (blue points) were derived from the best fit. The size of every data point reflects the size of the planetary radius.}
\label{MP_rel}
\end{figure}

\begin{table}
\caption{16 SWEEPS planet masses in units of Jupiter mass, estimated with the M-P anticorrelation. Names of the SWEEPS planets are indicated by number.}       
\label{tableA1}   
\centering             
\begin{tabular}{c c}    
\hline         
Name & M (M$_\textrm{J}$) \\  
\hline            
1  & 1.37       \\
2  & 1.52       \\
3  & 1.44       \\
4  & 0.76       \\
5  & 1.20       \\
6  & 1.03       \\
7  & 1.33       \\
8  & 1.53       \\
9  & 1.36       \\
10  & 1.64       \\
11  & 1.32       \\
12  & 1.05       \\
13  & 1.35       \\
14  & 1.05       \\
15  & 1.61       \\
16  & 1.51       \\
\hline
\end{tabular}
\end{table}

There exists a mass-period (M-P) relation for hot Jupiters. When we plot the hot-Jupiter masses versus their orbital period \citep[see also][]{mazeh2005, henning2018}, the anticorrelation (correlation coefficient: $-0.45$) between these two parameters is obvious. In Figure \ref{MP_rel} we show the M-P anticorrelation for 312 hot Jupiters from the EPE database (maroon points), including the masses of all 16 SWEEPS planets (cyan points) estimated from the best-fit line (black dashed) $M_\textrm{p} = -0.23 P + 1.73$. We list the estimated SWEEPS planet masses in Table \ref{tableA1}.

Comparing the masses of SWEEPS 4 and 11 obtained with the M-P relation with their masses in Table \ref{table1}, we observe that this empirical relation is not an accurate mass estimate for these planets. We should therefore be cautious with assuming the masses from Table \ref{table1} for the remaining 14 SWEEPS planets. However, because we lack other options, we used these masses to calculate the atmospheric erosion of the SWEEPS planets due to Sgr A*.

\subsection{Ultra-short hot Jupiters}

It is now understood that an orbit with $P < 1$ \textrm{day} around the parent star will result in a Roche-lobe flow of material. There exists a universal relation between the orbital period $P_{\textrm{orb}}$, mass $M_{\textrm{p}}$ and radius $R_{\textrm{p}}$ (P-M-R) of the donor planet of mass lower than the accreting star \citep{howell2001},

\begin{equation} \label{eq:mass_period_radius}
P_{\textrm{orb}} = 0.4 \left(\frac{M_{\textrm{p}}}{M_\textrm{J}}\right)^{-1/2} \left(\frac{R_{\textrm{p}}}{R_\textrm{J}}\right) \quad \textrm{days},
\end{equation}

where $R_\mathrm{p}$ and $M_\mathrm{p}$ are the planet radius and mass, and $M_\textrm{J}$ and $R_\textrm{J}$ are the Jupiter mass and radius, respectively.

\begin{table}
\caption{Ultra-short SWEEPS planet masses in units of Jupiter mass, estimated with Equation 4.4. Names of the SWEEPS planets are indicated by number.}       
\label{tableA2}   
\centering             
\begin{tabular}{c c}    
\hline        
Name & M (M$_\textrm{J}$) \\
\hline
2  & 0.878       \\
8  & 0.355       \\
10  & 3.01        \\
15  & 2.5        \\
16  & 0.83        \\
\hline                  
\end{tabular}
\end{table}

We employed this method to estimate the masses of the ultra-short SWEEPS planets (2, 8, 10, 15, and 16). This might increase the accuracy of our mass-loss estimation for these five planets. In Table \ref{tableA2} we list the masses of SWEEPS 2, 8, 10, 15, and 16, indicated by the corresponding number in the table, as estimated with the above relation.

\section{Distance on cosmological scales}

The luminosity distance $D_\textrm{L}$ essentially defines the actual distance traveled by photons from the object to the observer because it accounts for the expansion of the universe. For an object with redshift $z$, $D_\textrm{L}$ is defined as

\begin{equation} \label{eq:lum_distance}
D_\textrm{L} = (1 + z)D_\textrm{m}(z),\end{equation}

where $D_\textrm{m}(z)$ is the comoving distance, which in a flat universe is defined as

\begin{equation} \label{eq:comoving_distance}
D_\textrm{m}(z) = D_\textrm{H} \int\limits_{0}^{z} \frac{\mathrm{d}z}{\sqrt{\Omega_\textrm{m}(1+z)^3 + \Omega_\Lambda}},
\end{equation}

where $\Omega_\textrm{m}$ and $\Omega_\Lambda$ are the matter and vacuum density parameters, respectively, and $D_\textrm{H} = \frac{c}{H_\textrm{0}}$ is the Hubble distance, where $H_\textrm{0}$ is the Hubble constant.

\section{Time dilation}

In an expanding universe, the cosmological timescales observed on Earth are no longer the same. To avoid inaccurate estimations, the time dilation due to to redshift $z$ must be considered,
\begin{equation} \label{eq:t_o}
t_\textrm{o} = s(1 + z)t_\textrm{e},
\end{equation}

where $s$ is the stretch factor, estimated at $\sim 1$, while $t_\textrm{o}$ and $t_\textrm{e}$ are the observed and emitted timescales, respectively \citep{goldhaber1996}. It is evident from Equation (\ref{eq:t_o}) that this effect becomes increasingly important at higher redshifts. The rest-frame timescale of $50$ \textrm{Myr} for an AGN at $z=0.5$ would be observed on Earth for a period longer by $25$ \textrm{Myr}. These AGNs contribute the least to the total $F_{\mathrm{XUV}}$ due to their distance, therefore including this time dilation would have a very weak effect on our result.

\end{appendix}

\end{document}